\documentclass{template/jfm}

\usepackage[english]{babel}
\usepackage[T1]{fontenc}
\usepackage[utf8]{inputenc}
\usepackage{graphicx}
\usepackage[outdir=figs/]{epstopdf}
\usepackage{epsfig}
\usepackage[section]{placeins}
\usepackage{float}
\usepackage{subfloat}
\ifx\pgfplothandlerjumpmarkmid\undefined
  \def\pgfplothandlerjumpmarkmid{123}
\fi
\usepackage{pgfplots}
\usepackage{caption}
\usepackage{subcaption}
\usepackage{morefloats}
\usepackage{amssymb}
\usepackage{amsmath}
\usepackage{bm}
\usepackage{esint}
\usepackage{hyperref}
\usepackage[nameinlink]{cleveref}
\crefname{section}{§}{§§}
\Crefname{section}{§}{§§}
\usepackage[section]{placeins}
\usepackage{cancel}
\newcommand{\mansort}[1]{} 
\newcommand{\appropto}{\mathrel{\vcenter{
  \offinterlineskip\halign{\hfil$##$\cr
    \propto\cr\noalign{\kern2pt}\sim\cr\noalign{\kern-2pt}}}}}

\shorttitle{Inter-scale Energy Exchange in the Presence of
  Inhomogeneity and Coherence} \shortauthor{F. Alves Portela,
  G. Papadakis and J. C. Vassilicos}

\title{The role of Coherent Structures and Inhomogeneity in Near-Field
  Inter-Scale Turbulent Energy Transfers}

\author{F. Alves Portela\aff{1}, G. Papadakis\aff{2} \and
  J. C. Vassilicos\aff{3}\corresp{\email{john-christos.vassilicos@centralelille.fr}}}

\affiliation{\aff{1} School of Engineering Sciences, University of
  Southampton, Southampton, SO17 1BJ, UK\\
\aff{2} Department of Aeronautics, Imperial College
  London, London SW7 2AZ, UK\\
\aff{3} Univ. Lille, CNRS, ONERA, Arts et M\'etiers
  ParisTech, Centrale Lille, FRE 2017 - LMFL - Laboratoire de
  M\'ecanique des fluides de Lille - Kamp\'e de Feriet, F-59000 Lille,
  France}

\begin{document}

\maketitle

\begin{abstract}
We use Direct Numerical Simulation (DNS) data to study inter-scale and
inter-space energy exchanges in the near-field of a turbulent wake of
a square prism in terms of a K\'arm\'an-Howarth-Monin-Hill (KHMH)
equation written for a triple decomposition of the velocity field
which takes into account the presence of quasi-periodic vortex
shedding coherent structures. Concentrating attention on the plane of
the mean flow and on the geometric centreline, we calculate
orientation-averages of every term in the KHMH equation. The
near-field considered here ranges between 2 and 8 times the width $d$
of the square prism and is very inhomogeneous and out of equilibrium
so that non-stationarity and inhomogeneity contributions to the KHMH
balance are dominant. The mean flow produces kinetic energy which
feeds the vortex shedding coherent structures. In turn, these coherent
structures transfer their energy to the stochastic turbulent
fluctuations over all length-scales $r$ from the Taylor length
$\lambda$ to $d$ and dominate spatial turbulent transport of
small-scale two-point stochastic turbulent fluctuations. The
orientation-averaged non-linear inter-scale transfer rate $\Pi^{a}$
which was found to be approximately independent of $r$ by
\citet{AlvesPortela2017} in the range $\lambda\le r \le 0.3d$ at a
distance $x_{1}=2d$ from the square prism requires an inter-scale
transfer contribution of coherent structures for this approximate
constancy. However, the near-constancy of $\Pi^{a}$ in the range
$\lambda\le r\le d$ at $x_{1}=8d$ which was also found by
\citet{AlvesPortela2017} is mostly attributable to stochastic
fluctuations. Even so, the proximity of $-\Pi^{a}$ to the turbulence
dissipation rate $\varepsilon$ in the range $\lambda\le r\le d$ at
$x_{1}=8d$ does require inter-scale transfer contributions of the
coherent structures. Spatial inhomogeneity also makes a direct and
distinct contribution to $\Pi^{a}$, and the constancy of
$-\Pi^{a}/\varepsilon$ close to $1$ would not have been possible
without it either in this near-field flow. Finally, the
pressure-velocity term is also an important contributor to the KHMH
balance in this near-field, particularly at scales $r$ larger than
about $0.4d$, and appears to correlate with the purely stochastic
non-linear inter-scale transfer rate when the orientation average is
lifted.

\end{abstract}


\section{Introduction}\label{sec:intro}

Coherent flow structures are present in most turbulent flows. Coherent
structures associated with vortex shedding, in particular, are clearly
present in turbulent wakes. One can expect these structures to have
some impact on a two-point energy balance which takes into account
both inter-scale and inter-space energy transfers. Such an energy
balance which can be applied to turbulent flows which are not
necessarily homogeneous and isotropic has already been used by various
authors to analyse turbulent flows starting with \citet{Marati2004}
who applied it to turbulent channel flow. This energy balance, first
derived by \citet{Hill2002} (but see also \citet{Duchon2000}), is
sometimes referred to as the K\'arm\'an-Howarth-Monin-Hill (KHMH)
equation because it fully generalises the K\'arm\'an-Howarth-Monin
equation (see \citet{Frisch1995}) which is limited to homogeneous and
to periodic turbulence. To our knowledge, there has been, to date,
only one study of such an energy balance in a boundary free turbulent
shear flow which takes account of coherent structures. This is the
study of \citet{Thiesset2014a} who derived a KHMH equation written for
a triple decomposition, where the coherent quasi-periodic part of the
fluctuating velocity field is explicitly treated in the analysis as
distinct from the stochastic turbulent
fluctuations. \citet{Thiesset2014a} applied their two-point equation
to a turbulent wake of a cylinder and concentrated attention at
downstream distances between $10d$ and $40d$, where $d$ is the
diameter of the cylinder. They found that the coherent structures
impose a forcing on the stochastic fluctuations and proposed an
analytical model which describes the energy content of such structures
in scale space.

The one other study of the KHMH equation in a planar turbulent wake is
that of \citet{AlvesPortela2017} who looked at inter-scale and
inter-space exchanges in the near wake of a square prism of side width
$d$ but did not consider the effects of vortex shedding coherent
structures. They found that $\Pi^a$, the rate at which turbulent
energy is transferred across scales when averaged over orientations in
the plane of the mean-flow (plane normal to coordinate $x_3$), is
roughly constant, and in fact close to the turbulence dissipation rate
$\varepsilon$, over a wide range of scales at a distance $8d$ from the
square prism. Their Direct Numerical Simulation (DNS) showed that this
is also true, albeit over a much reduced range of length-scales, at a
distance $2d$ from the square prism. Their KHMH analysis made it clear
that this Kolmogorov-sounding constancy of $\Pi^a$ cannot be the
result of a Kolmogorov equilibrium cascade given that the near-field
region of the flow where it is observed is very inhomogeneous,
anisotropic and out of equilibrium. One is therefore naturally faced
with the question of the role of the coherent structures in
establishing $-\Pi^a/\varepsilon \approx 1$ and the extent in which
this approximate constancy is due to the stochastic component of the
turbulent fluctuations. We also attempt to address the direct
contribution of spatial inhomogeneity to the behaviour of $\Pi^a$.


In this paper we use the triple decomposition KHMH equations of
\citet{Thiesset2014a} which we slightly generalise to include mean
flow velocity differences. We analyse the data obtained by
\citet{AlvesPortela2017} from their DNS of the turbulent planar wake
of a square prism of side length $d$. The inlet free-stream velocity
$U_{\infty}$ in this DNS is such that $U_{\infty}d/\nu = 3900$ where
$\nu$ is the fluid's kinematic viscosity. We refer to
\citet{AlvesPortela2017} for details of this DNS.

In \cref{sec:tripdecomp} we explain how the triple decomposition is
carried out and how we extract from the time-varying fields of
velocity and pressure a contribution associated with the vortex
shedding.  In \cref{sec:cascadetripdecomp} we detail the
scale-by-scale KHMH budgets that we use in this paper to explore
combined inter-scale and inter-space transfers in the near wake of a
square prism and in \cref{sec:orientation-averaged} we report on the
various terms in our KHMH budgets in an orientation-averaged sense.
\cref{subsec:tripledecomp_constPi} presents our results on inter-scale
energy transfers and scale-space fluxes and we conclude in
\cref{sec:conclusion}.


\section{Triply Decomposed Velocity Field}\label{sec:tripdecomp}

The Reynolds decomposition distinguishes between the mean field and
the fluctuating field. When the flow exhibits a well-defined
non-stochastic (e.g. periodic) flow feature, one can further decompose
the fluctuating field into a coherent field and a stochastic field
\citep[][]{Reynolds1972,Hussain1970}. The velocity field is therefore
the sum of three fields: ${\bf{u}_{full}} = \bf{U} + \tilde{\bf{u}} +
\bf{u}'$ where $\bf{U}$ is the mean velocity field obtained by
time-averaging ${\bf{u}_{full}}$, and where $\tilde{\bf{u}}$ and
$\bf{u}'$ are the coherent and stochastic parts, respectively, of the
fluctuating velocity field. The coherent fluctuating velocity
$\tilde{\bf{u}}$ is obtained by phase-averaging ${\bf{u}_{full}} -
\bf{U}$ and the stochastic fluctuating velocity is the remainder and
is obtained from $\bf{u}' = {\bf{u}_{full}} -\bf{U}
-\tilde{\bf{u}}$. If ${\bf{u}_{full}}$ is incompressible, $\bf{U}$,
$\tilde{\bf{u}}$ and $\bf{u}'$ are incompressible too. With similar
notation one also decomposes the pressure field: $p = P + \tilde{p} +
p'$. In the present work which is concerned with the planar wake of a
square prism, both time- and phase-averaging operations also involve
averaging in the span-wise direction, i.e. in the direction $x_{3}$
which is normal to the plane of the average wake flow. Fluid
velocities and spatial coordinates in the stream-wise direction are
denoted by $U_1$, $\tilde{u}_{1}$, $u'_{1}$, and $x_1$ respectively;
in the cross-stream direction they are $U_2$, $\tilde{u}_{2}$,
$u'_{2}$, and $x_2$. The span-wise fluid velocity components are $U_3$,
$\tilde{u}_{3}$, $u'_{3}$.

The definitions of $\tilde{\bf{u}}$ and $\tilde{p}$ require a
reference phase. One can obtain a reference phase from a pressure tap
on the cylinder \citep[see e.g.][]{Braza2006} or from the fluctuating
velocity signal, either from within the turbulent flow after
appropriately filtering \citep[see e.g.][]{Thiesset2014a} or from the
outside of the turbulent core \citep[see
  e.g.][]{Davies1976}. \citet{Wlezien1979} provide an extensive
comparison of different methods with focus on experimental techniques.

In the present analysis, the phase angle $\phi$ used to compute
phase-averages is extracted from the Hilbert transform of the lift
coefficient $C_L$ \citep[see][for details on the Hilbert
  transform]{Feldman2011}. This choice follows naturally from the fact
that the lift on the square prism in our flow closely follows a
sinusoid in time.

\begin{figure}
\centering
\includegraphics{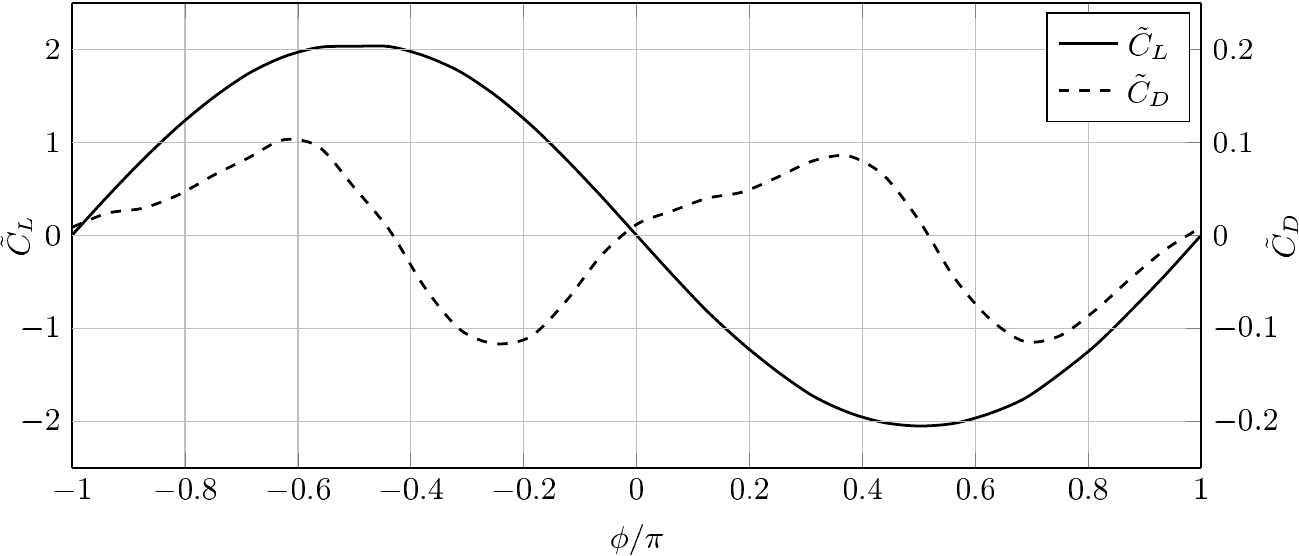}
\caption{Evolution of phase-averaged lift and drag coefficients
  $\tilde{C}_L$ and $\tilde{C}_D$ along the normalised phase
  $\phi/\pi$.}\label{fig:phaseclcd}
\end{figure}

The data being discrete in time, $\phi$ was binned into $32$ groups. A
smaller bin size would have improved phase-resolution but would have
reduced statistical convergence (as fewer samples would have fallen
within each bin). Thus, each time instant is associated with a value
$\phi=-\pi+n\frac{2\pi}{32}$ where $0<n<31$. The resulting
phase-averaged lift and drag coefficients are plotted in
\cref{fig:phaseclcd} versus the phase angle $\phi$, where $\phi=0$ has
been chosen such that $\tilde{C}_L(\phi=0)=0$.

\begin{figure}
\centering
    \begin{subfigure}[t]{0.5\textwidth}
        \centering
        \includegraphics[width=\textwidth]{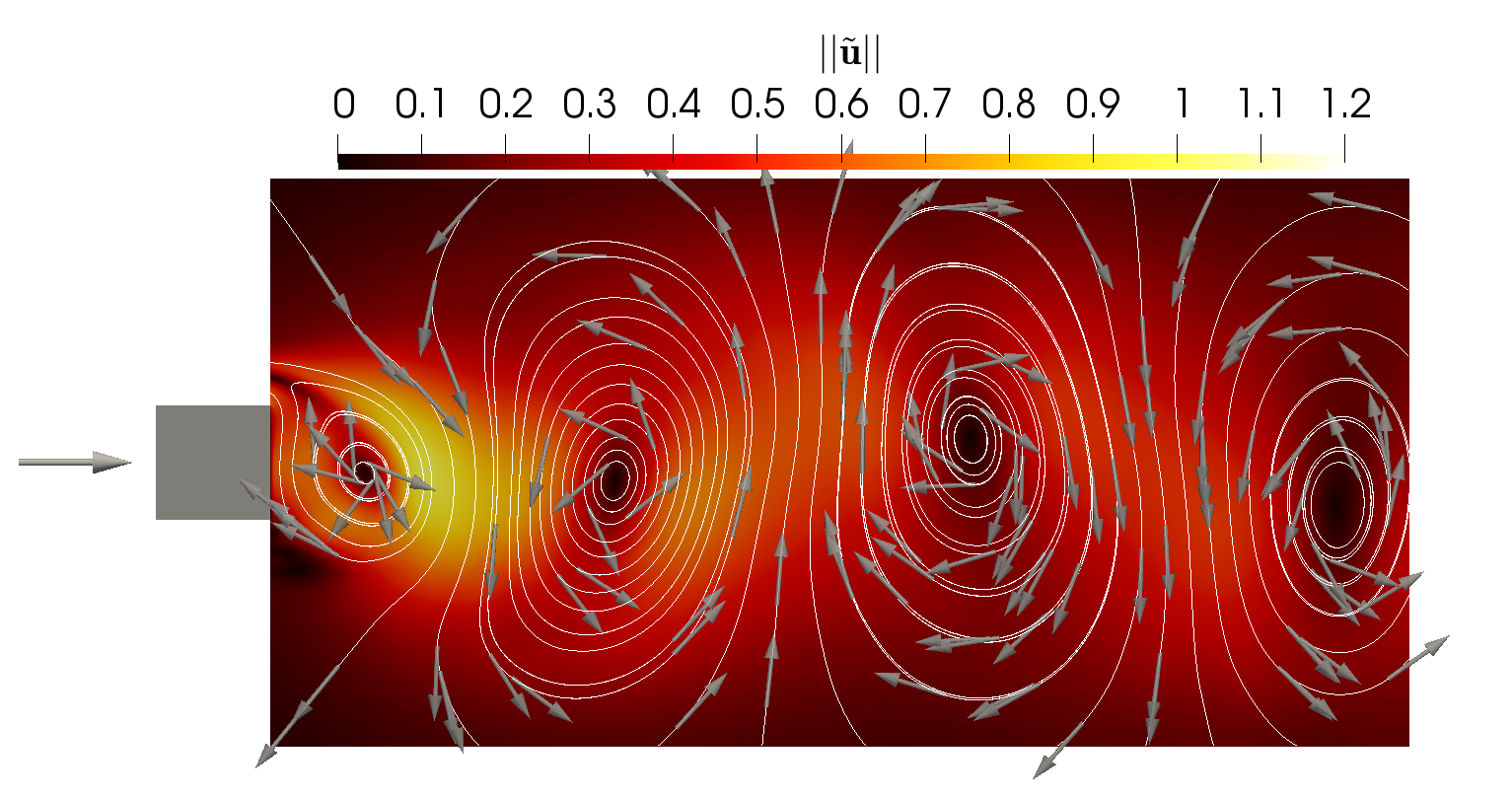}
    \end{subfigure}%
    ~
    \begin{subfigure}[t]{0.5\textwidth}
\centering
        \includegraphics[width=\textwidth]{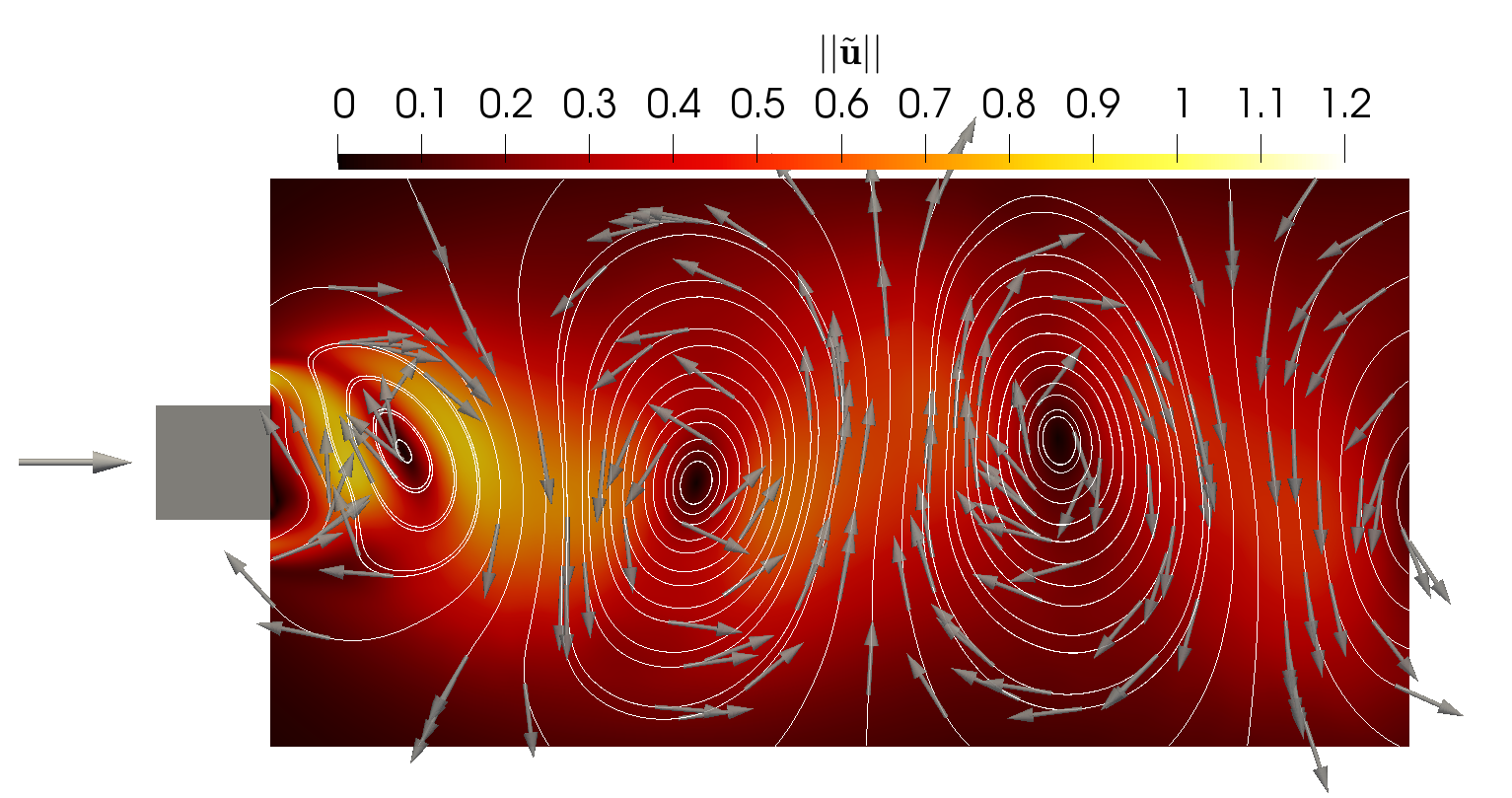}
    \end{subfigure}
    \\
    \begin{subfigure}[t]{0.5\textwidth}
        \centering
        \includegraphics[width=\textwidth]{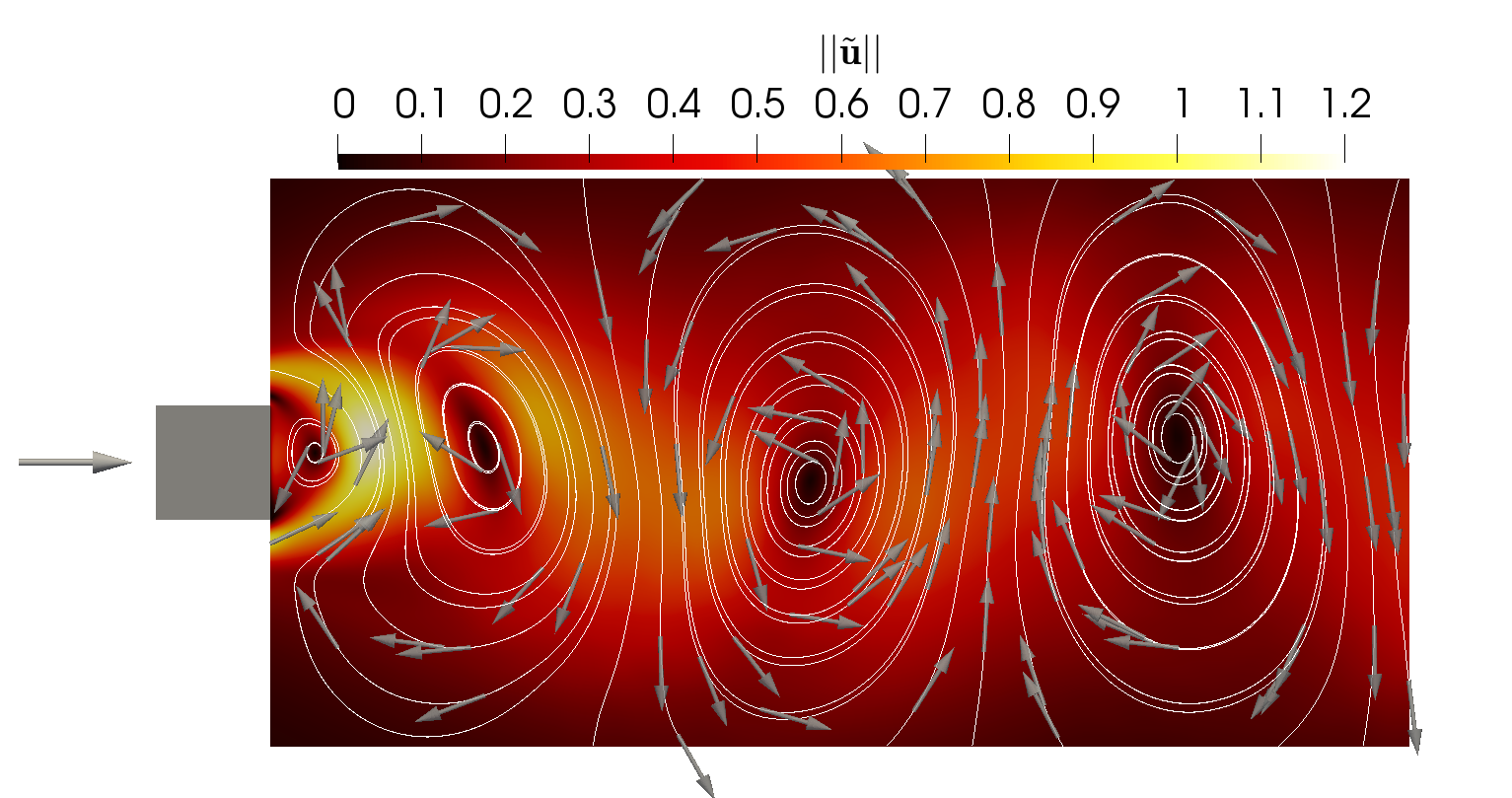}
    \end{subfigure}%
    ~
    \begin{subfigure}[t]{0.5\textwidth}
\centering
        \includegraphics[width=\textwidth]{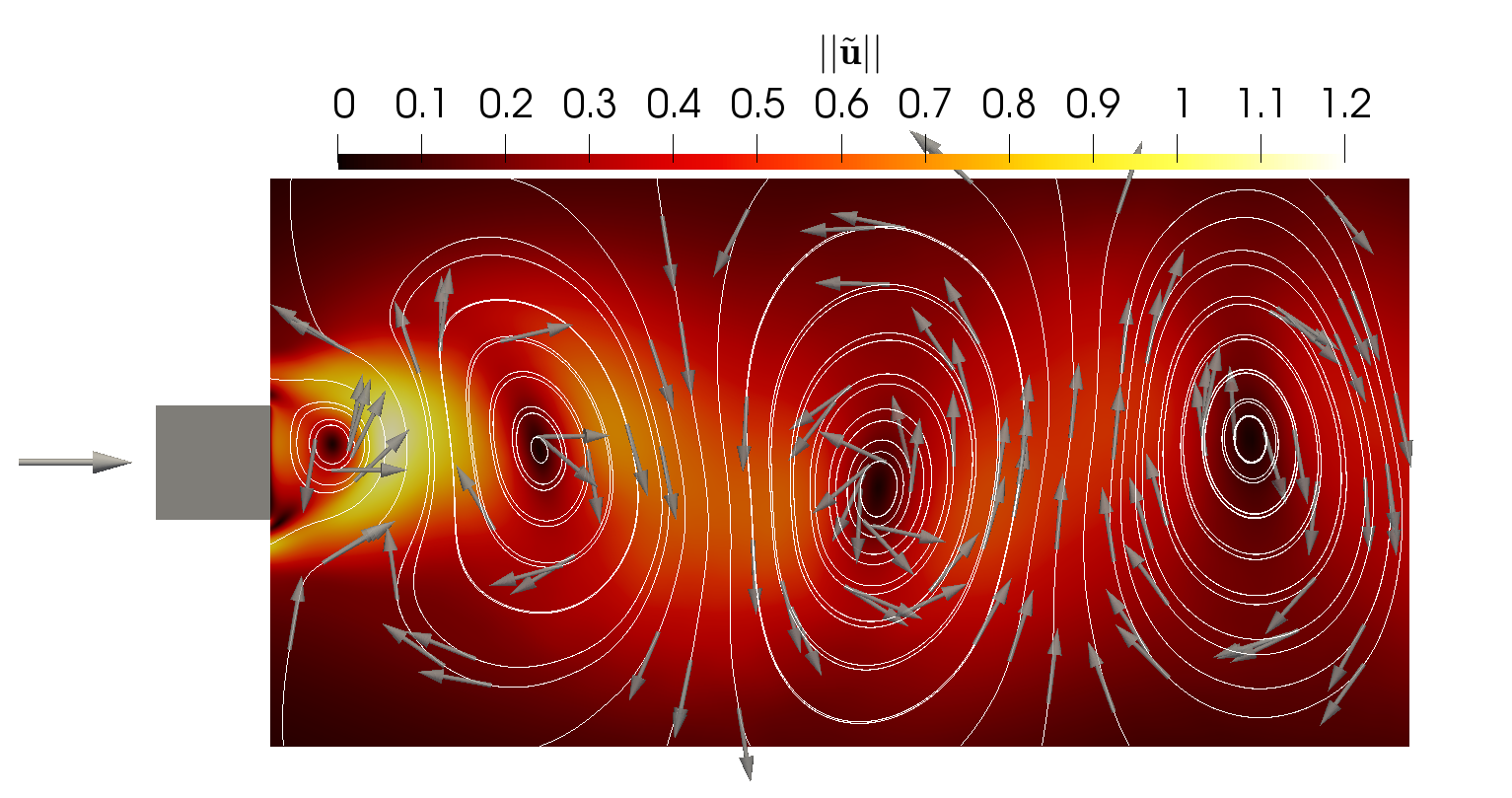}
    \end{subfigure}
\caption{Contours of the magnitude of the phase-averaged velocity
  $\tilde{\mathbf{u}}$ (normalised by $U_\infty$), and white unit
  vectors locally parallel to $\tilde{\mathbf{u}}$. The large arrow on
  the left indicates the direction of the free-stream velocity
  $U_\infty$. Using the phase angles shown in \cref{fig:phaseclcd}, on
  the top row $\phi=0$ (left) and $\phi=\frac{1}{4}\pi$ (right); on
  the bottom row $\phi=\frac{1}{2}\pi$ (left) and
  $\phi=\frac{3}{4}\pi$ (right).}\label{fig:mag_utilde_phiN}
\end{figure}

The phase-averaged velocity field $\tilde{\mathbf{u}}$ is shown in
\cref{fig:mag_utilde_phiN} for four different values of $\phi$: $0$,
$\frac{1}{4}\pi$, $\frac{1}{2}\pi$ and $\frac{3}{4}\pi$.

It clearly displays a structure similar to that of the von
K\'{a}rm\'{a}n vortex street where the alternating vortices display
opposite circulation, the positive ones travelling slightly above and
the negative ones slightly below the centreline. Note that
$\tilde{u}_{3} = 0$ uniformly and that $\tilde{u}_{1}$ and
$\tilde{u}_{2}$ depend on $x_1$ and $x_2$ but not on $x_3$.

The coherent vorticity field ${\bf \nabla} \times \tilde{\bf{u}}$ is
aligned with the span-wise direction and therefore has only one
non-zero component $\tilde{\omega}_{3}$. As shown in
\citet{AlvesPortela2018a} for this exact same flow (see their fig. 3),
lines of constant vorticity approximately coincide with streamlines of
$\tilde{\bf{u}}$. As discussed in \citet{Hussain1983}, the streamlines
are not necessarily good indicators of the presence of coherent
structures, but \citet{Lyn1995} argue that, apart from the base region
in the very near wake where the coherent structures are formed, there
is indeed a correspondence between iso-vorticity and streamlines in
identifying coherent structures.

\begin{figure}
    \centering
    \begin{subfigure}[t]{0.4\textwidth}
        \centering
                \includegraphics[width=\textwidth]{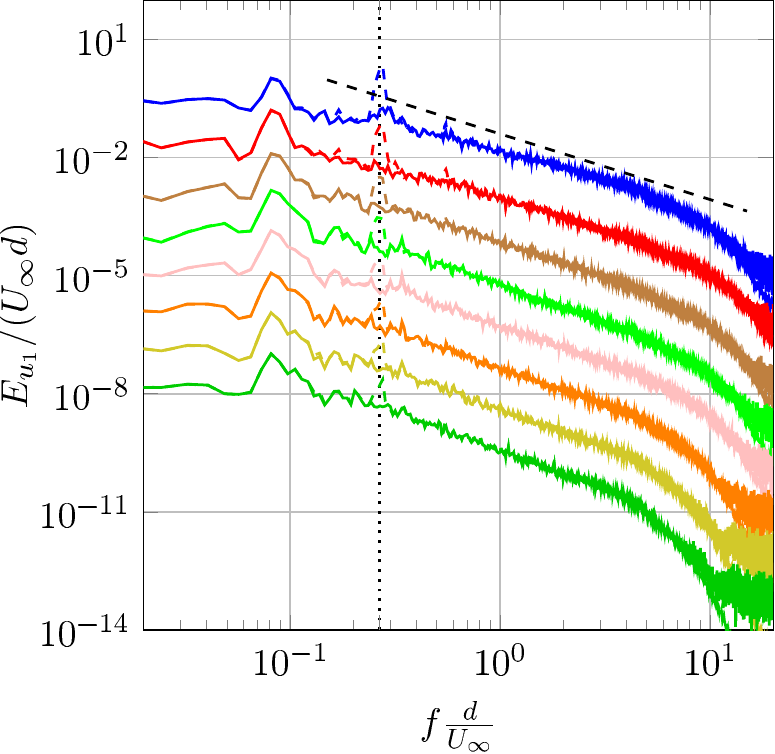}
    \end{subfigure}%
    ~
    \begin{subfigure}[t]{0.4\textwidth}
\centering
         \includegraphics[width=\textwidth]{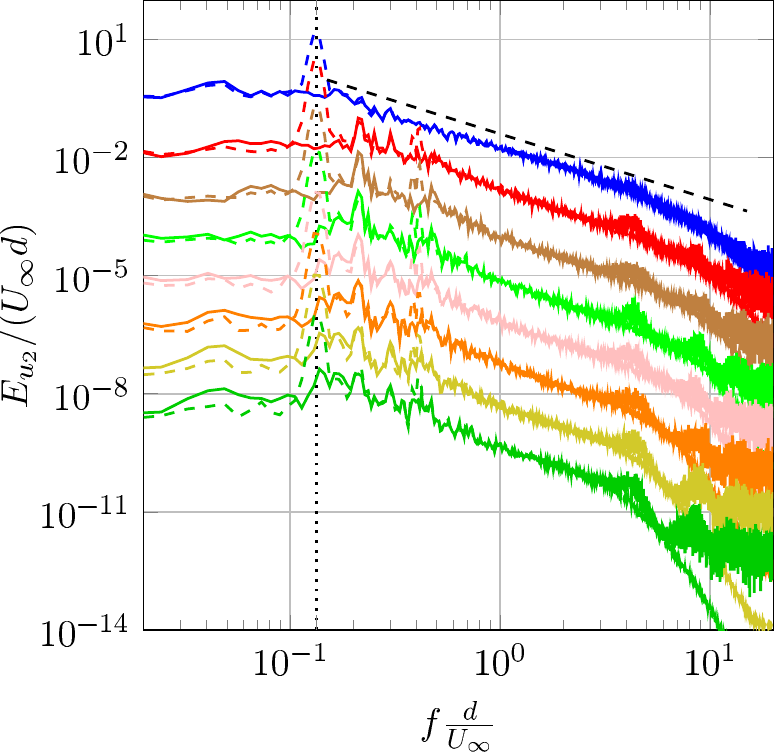}
    \end{subfigure}
    \caption{Power spectrum densities normalised by $ U_{\infty} d$ of
      stream-wise (left) and cross-stream (right) fluctuating
      velocities, before (dashed lines) and after (full lines)
      removing the phase component, between $x_1/d=1$ (blue/top) and
      $x_1/d=8$ (dark green/bottom) offset by one decade every
      $d$. The dashed line indicates a slope of $-5/3$ and the dotted
      line indicates $f=2f_s$ (left) and $f=f_s$
      (right).}\label{fig:Euv_phasestoch}
\end{figure}

The spectra of the full fluctuating velocity component $\tilde{u}_1 +
u'_1$ and $\tilde{u}_2 + u'_2$ are compared to those of their
stochastic counterparts $u'_1$ and $u'_2$ in
\cref{fig:Euv_phasestoch}. As is well known, the shedding frequency is
double in the spectrum of $\tilde{u}_1 + u'_1$ compared to the
spectrum of $\tilde{u}_2 + u'_2$, and we checked that it corresponds
to the distance between coherent vortices in
\cref{fig:mag_utilde_phiN} (the distance between successive such
vortices does not vary much). Note that the energetic peak present at
the shedding frequency in the spectrum of $\tilde{u}_1 + u'_1$ is
absent in the spectrum of $u'_1$ and that the energetic peak present
at the shedding frequency in the spectrum of $\tilde{u}_2 + u'_2$ is
absent in the spectrum of $u'_2$.

\begin{figure}
\centering
\includegraphics{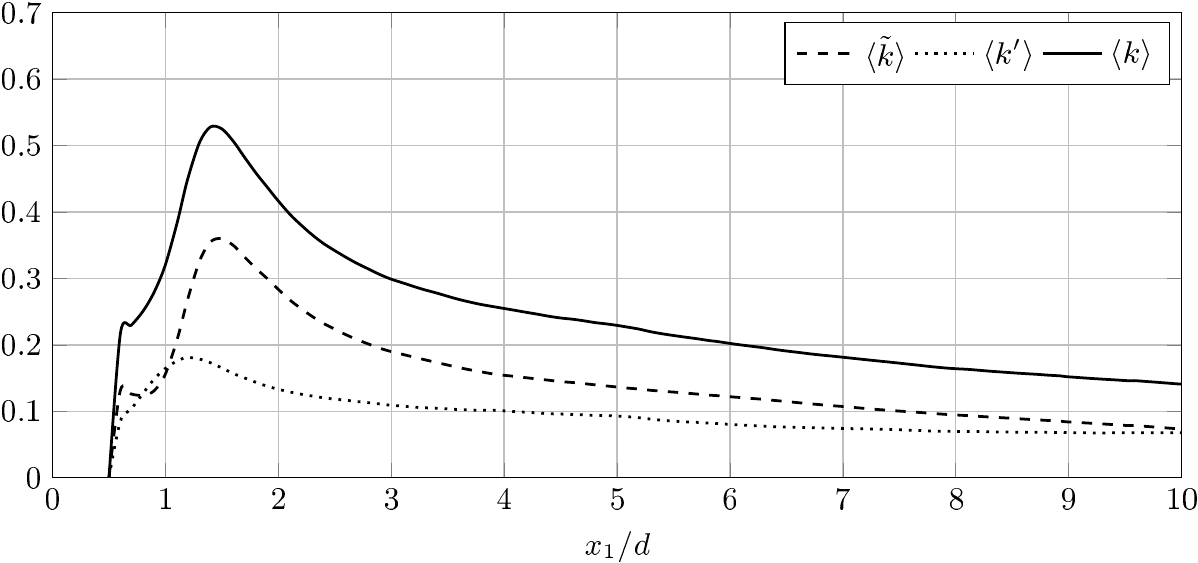}
\caption{Profiles of kinetic energies $\tilde{k}$ and $k'$ computed
  from the phase and stochastic components, respectively, along the
  centreline. The total kinetic energy $k=\tilde{k}+k'$ is also shown
  for comparison.}\label{fig:tke_phase_centreline}
\end{figure}

In conclusion, the phase-averaged fluctuating velocity
$\tilde{\mathbf{u}}$ is representative of the coherent structures
in the present flow as it contains the shedding's characteristic time
signature, and its spatial distribution (\cref{fig:mag_utilde_phiN})
is one of approximately periodic large scale vortices.

In \citet{Hussain1983,Hussain1987} it is argued that these coherent
structures do not necessarily provide a large contribution to the
turbulent kinetic energy. Of course, the regions of the flow
considered by these authors are much further downstream than the
region of the flow studied here. \Cref{fig:tke_phase_centreline} makes
it clear that the coherent structures contribute most of the turbulent
kinetic energy $k\equiv {1\over 2}\langle \vert \tilde{\mathbf{u}}
+\mathbf{u}'\vert^{2} \rangle$ in the near wake considered here and
that their contribution ($\tilde{k} \equiv {1\over 2} \langle \vert
\tilde{\mathbf{u}}\vert^{2} \rangle$) decreases, in the direction of
the mean flow, at a faster rate than the kinetic energy associated
with the stochastic motions ($k' \equiv {1\over 2} \langle \vert
\mathbf{u}'\vert^{2} \rangle$) in-line with
\citet{Hussain1983,Hussain1987}. (The brackets $\langle ... \rangle$ symbolise
combined time- and span-wise-average operations using approximately
$10^3$ snapshots spanning just over $32$ shedding cycles. The
additional span-wise average involves $150$ planes in the span-wise
direction which is statistically homogeneous. This level of statistics
proved sufficient to converge the averages of all the quantities
presented in this paper.) Note that $k = \tilde{k} +k'$. Note also
that the Taylor length-based Reynolds number $Re_{\lambda}$ varies on
the centreline from about 120 at $x_{1}/d =2$ to about 170 at $x_{1}/d
=10$ if it is defined on the basis of $\sqrt{\langle {u'_{1}}^{2} \rangle}$ and from
about 100 at $x_{1}/d =2$ to about 190 at $x_{1}/d =10$ if it is
defined on the basis of $\sqrt{2k'/3}$.

 In the following section we introduce scale-by-scale energy budgets
 adapted to the triple decomposition of a velocity field into its mean
 and its coherent and stochastic fluctuations.

\section{Scale-by-scale Energy Budgets}\label{sec:cascadetripdecomp}

The most general forms of scale-by-scale energy budget have been
derived by \citet{Hill1997,Hill2001,Hill2002a} and \citet{Duchon2000}
without making any assumption on the nature of the turbulence. Using
the Reynolds decomposition and averaging over time in general but also
in the span-wise direction for this paper's particular flow, this
equation (which we refer to as K\'arm\'an-Howarth-Monin-Hill (KHMH)
equation) follows from the Navier-Stokes equation and
incompressibility and takes the form

\begin{multline}\label{eq:KHMH}
\frac{U_i^++U_i^-}{2} {\frac{\partial \langle\delta
    q^2\rangle}{\partial x_i}}+{ \frac{\partial \langle\delta u_i
    \delta q^2\rangle}{\partial r_i}}+{\frac{ \partial \delta U_i
    \langle\delta q^2\rangle}{\partial r_i}}=-2\langle{\delta u_i
  \delta u_j}\rangle\frac{\partial \delta U_j}{\partial
  r_i}-\\-\langle{\left(u_i^++u_i^-\right)\delta
  u_j}\rangle\frac{\partial \delta U_j}{\partial x_i}-{\frac{\partial
    \langle \frac{u_i^++u_i^-}{2} \delta q^2\rangle}{\partial x_i}}- 2
\frac{\partial \langle{\delta u_i \delta p \rangle}}{\partial x_i} +
\nu \frac{1}{2}\frac{\partial^2 \langle{\delta q^2}\rangle}{\partial
  x_i \partial x_i} + \\ + 2 \nu \frac{\partial^2 \langle{\delta
    q^2}\rangle}{\partial r_i \partial r_i} - 4\nu
\left(\langle{\frac{\partial \delta u_j}{\partial x_i}\frac{\partial
    \delta u_j}{\partial
    x_i}}\rangle+\frac{1}{4}\langle{\frac{\partial \delta
    u_j}{\partial r_i}\frac{\partial \delta u_j}{\partial
    r_i}}\rangle\right)
\end{multline}
where $\delta q^{2} \equiv \delta u_{i} \delta u_{i}$ in terms of the
fluctuating velocity differences $\delta u_i \equiv (\tilde{u}_i^+ +
{u'_i}^{+}) -(\tilde{u}_i^- + {u'_i}^{-})$ (for components $i=1,2,3$),
$\delta U_{i} \equiv U_i^+-U_i^-$, $\delta p \equiv (\tilde{p}^+ +
p'^{+}) -(\tilde{p}^- +p'^{-})$, and the superscripts $+$ and $-$
distinguish quantities evaluated at positions ${\bf \xi}^{+} \equiv
{\bf x}+{\bf r}/2$ and ${\bf \xi}^{-} \equiv {\bf x}-{\bf r}/2$,
respectively; e.g. $u^{+}_{i}\equiv \tilde{u}_i^+ + {u'_i}^+$ and
$u^{-}_{i}\equiv \tilde{u}_i^- + {u'_i}^-$ are the full fluctuating
velocity components at ${{\bf \xi}^{+}}$ and ${{\bf \xi}^{-}}$
respectively. Equation (\ref{eq:KHMH}) is written in a six-dimensional
reference frame $x_i,r_i$ where coordinates $x_i$ of ${\bf x}$ are
associated with a location in physical space and the scale space is
the space of all separations and orientations ${\bf r} = (r_{1},
r_{2}, r_{3})$ between two-points (we refer to $r= \vert {\bf r}
\vert$ as a scale).  If the average operation is not over time but
over realisations, then the extra term $\frac{\partial \langle\delta
  q^2\rangle}{\partial t}$ can also be present on the left hand side
of equation (\ref{eq:KHMH}). (Note that an even more general form of
the KHMH equation can be obtained without any decomposition and
without any averaging operation, see \citet{Duchon2000},
\citet{Hill2002a} and \citet{Yasuda2018}.)

Following \citet{Valente2013,Gomes-Fernandes2015, AlvesPortela2017},
each term in (\ref{eq:KHMH}), re-written as
\begin{equation}\label{eq:KHMHsimple}
\mathcal{A} + \Pi + \Pi_U = \mathcal{P} + \mathcal{T}_u +
\mathcal{T}_p + \mathcal{D}_x + \mathcal{D}_r - \varepsilon_r , 
\end{equation}
is associated with a physical process in the budget of $\langle\delta
q^2\rangle$ as follows:
\begin{itemize}

   \item $4 \mathcal{A} = \frac{U_i^++U_i^-}{2} {\frac{\partial
       \langle\delta q^2\rangle}{\partial x_i}} $ is the mean
     advection term.

   \item $4 \Pi = \frac{\partial \langle\delta u_i \delta
     q^2\rangle}{\partial r_i}$ is the non-linear inter-scale transfer
     rate which accounts for the effect of non-linear interactions in
     redistributing $\delta q^2 $ within the $r_i$ space and is given
     by the divergence in scale space of the flux $\langle\delta u_i
     \delta q^2\rangle$.

     \item $4 \Pi_U = \frac{ \partial \delta U_i \langle\delta
       q^2\rangle}{\partial r_i}$ is the linear inter-scale transfer
       rate. (The term ``linear'' used here does not mean that a
       linearisation of the Navier-Stokes equation has been
       performed.)

    \item $4 \mathcal{P} = -2\langle{\delta u_i \delta
      u_j}\rangle\frac{\partial \delta U_j}{\partial
      r_i}-\langle{\left(u_i^++u_i^-\right)\delta
      u_j}\rangle\frac{\partial \delta U_j}{\partial x_i}$ can be
      associated with the production of $\langle \delta q^2 \rangle$
      by mean flow gradients. (See \citet{AlvesPortela2017} for more
      details.)

  \item $4 \mathcal{T}_u = -\frac{\partial \langle
    \frac{u_i^++u_i^-}{2} \delta q^2\rangle}{\partial x_i}$ is the
    transport of $\delta q^2$ in physical space due to turbulent
    fluctuations.

    \item $4 \mathcal{T}_p = -2 \frac{\partial \langle{\delta u_i
        \delta p \rangle}}{\partial x_i}$ is the pressure-velocity
      term, equal to $-2$ times the correlation between fluctuating
      velocity differences and differences of fluctuating pressure
      gradient.

      \item $4 \mathcal{D}_x = \nu \frac{1}{2}\frac{\partial^2
        \langle{\delta q^2}\rangle}{\partial x_i \partial x_i}$ is the
        viscous diffusion in physical space.

        \item $4 \mathcal{D}_r = 2 \nu \frac{\partial^2 \langle{\delta
            q^2}\rangle}{\partial r_i\partial r_i}$ is the viscous
          diffusion in scale space. This term is equal to the
          dissipation $\varepsilon$ when the two points coincide
          ($r=0$) and can be shown \citep[see Appendix B
            in][]{Valente2013} to be negligible for separations larger
          than the Taylor micro-scale.

     \item $4\varepsilon_r = 4\nu \left(\langle{\frac{\partial \delta
         u_j}{\partial x_i}\frac{\partial \delta u_j}{\partial
         x_i}}\rangle+\frac{1}{4}\langle{\frac{\partial \delta
         u_j}{\partial r_i}\frac{\partial \delta u_j}{\partial
         r_i}}\rangle\right)$ and $\varepsilon_r$ is actually the
       two-point average dissipation rate $\varepsilon_r =
       \frac{\varepsilon^++\varepsilon^-}{2}$ as it equals
       $\frac{1}{2}\nu \left(\langle{\frac{\partial u_j^+}{\partial
           x_i^+}\frac{\partial u_j^+}{\partial
           x_i^+}}\rangle+\langle{\frac{\partial u_j^-}{\partial
           x_i^-}\frac{\partial u_j^-}{\partial
           x_i^-}}\rangle\right)$.

\end{itemize}
       
With the triple decomposition introduced in \cref{sec:tripdecomp} one
can decompose the second order structure function $\langle \delta
{q}^2 \rangle$ into a stochastic and coherent part, i.e. $\langle
\delta {q}^2 \rangle = \langle \delta \tilde{q}^2\rangle +
\langle\delta {q'}^2\rangle$ where $\delta \tilde{q}^2 \equiv \delta
\tilde{u}_{i} \delta \tilde{u}_{i}$ with $\delta \tilde{u}_{i}\equiv
\tilde{u}_i^+ -\tilde{u}_i^-$ and $\delta {q'}^2 \equiv \delta u'_{i}
\delta u'_{i}$ with $\delta u'_{i} \equiv {u'_i}^+ - {u'_i}^-$.  The
fluctuating pressure difference $\delta p$ is also decomposed in a
similar way, i.e. $\delta p = \delta \tilde{p} + \delta p'$ where
$\delta \tilde{p} \equiv \tilde{p}^+ -\tilde{p}^-$ and $\delta p'
\equiv p'^+ -p'^-$.

This decomposition into stochastic and coherent fluctuations warrants
new scale-by-scale energy budgets to be derived and this was done by
\citet{Thiesset2014a} by neglecting mean flow velocity differences
$\delta U_{i}$. The resulting slightly more general equations for
$\langle\delta {q'}^2\rangle$ and $\langle\delta \tilde{q}^2\rangle$
without neglecting $\delta U_{i}$ are, respectively,
\begin{multline}\label{eq:tripleKHM_stoch}
\frac{U_i^++U_i^-}{2} \frac{\partial}{\partial x_i}\langle \delta
     {q'}^2 \rangle + \frac{\partial}{\partial r_i}\langle\delta u'_i
     \delta {q'}^2\rangle + \frac{\partial}{\partial r_i}\langle\delta
     \tilde{u}_i \delta {q'}^2\rangle + \frac{\partial}{\partial
       r_i}\langle\delta U_i \delta {q'}^2 \rangle= \\ -\langle\delta
     u'_i ({u'_j}^++{u'_j}^-) \rangle \frac{\partial \delta
       U_i}{\partial x_j} -2 \langle \delta u'_i \delta u'_j \rangle
     \frac{\partial \delta U_i}{\partial r_j} - \langle\delta u'_i
     ({u'_j}^++{u'_j}^-) \frac{\partial \delta \tilde{u}_i}{\partial
       x_j} \rangle-2\langle \delta {u'_i} \delta {u'_j}
     \frac{\partial \delta \tilde{u}_i}{\partial r_j} \rangle \\-
     \langle\frac{\tilde{u}_i^++\tilde{u}_i^-}{2}\frac{\partial \delta
       {q'}^2}{\partial x_i}\rangle
     -\langle\frac{{u'_i}^++{u'_i}^-}{2}\frac{\partial \delta
       {q'}^2}{\partial x_i}\rangle - 2\langle \delta u'_i \delta
     \frac{\partial p'}{\partial x_i}\rangle \\+\nu\langle
     \frac{1}{2}\frac{\partial^2 \delta {q'}^2}{\partial x_j \partial
       x_j}\rangle + 2\nu \langle \frac{\partial^2 \delta
       {q'}^2}{\partial r_j \partial r_j}\rangle- 4\nu
     \left(\langle{\frac{\partial \delta u'_j}{\partial
         x_i}\frac{\partial \delta u'_j}{\partial
         x_i}}\rangle+\frac{1}{4}\langle{\frac{\partial \delta
         u'_j}{\partial r_i}\frac{\partial \delta u'_j}{\partial
         r_i}}\rangle\right)
\end{multline}
and
\begin{multline}\label{eq:tripleKHM_phase}
\phantom{.} \frac{U_i^++U_i^-}{2}\frac{\partial}{\partial x_i} \langle
\delta \tilde{q}^2 \rangle + \frac{\partial}{\partial
  r_i}\langle\delta \tilde{u}_i \delta \tilde{q}^2 \rangle + 2
\frac{\partial}{\partial r_i} \langle \delta u'_i \delta u'_j \delta
\tilde{u}_j\rangle + \frac{\partial}{\partial r_i}\langle\delta U_i
\delta \tilde{q}^2\rangle = \\ -\langle\delta \tilde{u}_i
(\tilde{u}_j^++\tilde{u}_j^-)\rangle \frac{\partial \delta
  U_i}{\partial x_j} -2\langle \delta \tilde{u}_i \delta \tilde{u}_j
\rangle \frac{\partial \delta U_i}{\partial r_j} + \langle\delta u'_i
({u'_j}^++{u'_j}^-) \frac{\partial \delta \tilde{u}_i}{\partial x_j}
\rangle+\langle 2\delta {u'_i} \delta {u'_j} \frac{\partial \delta
  \tilde{u}_i}{\partial r_j} \rangle
\\ -\langle\frac{\tilde{u}_i^++\tilde{u}_i^-}{2}\frac{\partial \delta
  \tilde{q}^2}{\partial x_i}\rangle -\langle\frac{\partial}{\partial
  x_i}[({u'_i}^++{u'_i}^-)\delta u'_j \delta
  \tilde{u}_j]\rangle-2\langle \delta \tilde{u}_i \delta
\frac{\partial \tilde{p}}{\partial x_i}\rangle \\ +\nu\langle
\frac{1}{2}\frac{\partial^2 \delta \tilde{q}^2}{\partial x_j \partial
  x_j}\rangle + 2\nu\langle \frac{\partial^2 \delta
  \tilde{q}^2}{\partial r_j \partial r_j}\rangle- 4\nu
\left(\langle{\frac{\partial \delta \tilde{u}_j}{\partial
    x_i}\frac{\partial \delta \tilde{u}_j}{\partial
    x_i}}\rangle+\frac{1}{4}\langle{\frac{\partial \delta
    \tilde{u}_j}{\partial r_i}\frac{\partial \delta
    \tilde{u}_j}{\partial r_i}}\rangle\right) .
\end{multline}

Evidently both \cref{eq:tripleKHM_stoch} and \cref{eq:tripleKHM_phase}
are rather similar to the KHMH \cref{eq:KHMH} and we therefore make
use of similar notation to identify the individual terms:
\begin{equation}\label{eq:tripleKHMsimple_stoch}
\mathcal{A}' + \Pi' + \Pi_{\tilde{u}}' + \Pi_U' = \mathcal{P}_U' +
\mathcal{P}_{\tilde{u}}' + \mathcal{T}_{\tilde{u}}' +
\mathcal{T}_{u'}' + \mathcal{T}_{p'}' + \mathcal{D}_x' +
\mathcal{D}_r' - \varepsilon'_r
\end{equation}
for \cref{eq:tripleKHM_stoch} and
\begin{equation}\label{eq:tripleKHMsimple_phase}
\mathcal{\tilde{A}} + \tilde{\Pi}_{\tilde{u}} +
\tilde{\Pi}_{\mathcal{P}_{\tilde{u}}} + \tilde{\Pi}_U =
\mathcal{\tilde{P}}_U - \mathcal{P}_{\tilde{u}}'
+\mathcal{\tilde{T}}_{\tilde{u}} +
\mathcal{\tilde{T}}_{\mathcal{P}_{\tilde{u}}} +
\mathcal{\tilde{T}}_{\tilde{p}} +
\mathcal{\tilde{D}}_x + \mathcal{\tilde{D}}_r - \tilde{\varepsilon}_r
\end{equation}
for \cref{eq:tripleKHM_phase}. $4\mathcal{A}'$, $4\Pi'$,
$4\Pi_{\tilde{u}}'$ and $4\Pi_U'$ correspond to the first, second,
third and fourth terms in the first line of \cref{eq:tripleKHM_stoch}
and $4\mathcal{\tilde{A}}$, $4\tilde{\Pi}_{\tilde{u}}$,
$4\tilde{\Pi}_{\mathcal{P}_{\tilde{u}}}$ and $4\tilde{\Pi}_U$
correspond to the first, second, third and fourth terms in the first
line of \cref{eq:tripleKHM_phase}. $4\mathcal{P}_U'$ and
$4\mathcal{\tilde{P}}_U$ correspond to the sum of the first and second
terms in the second line of \cref{eq:tripleKHM_stoch} and
\cref{eq:tripleKHM_phase} respectively. For the same reasons given for
$4\mathcal{P}$ by \citet{AlvesPortela2017}, $4\mathcal{P}_U'$ and
$4\mathcal{\tilde{P}}_U$ are production terms of $\langle \delta
q'^{2}\rangle$ and $\langle \delta \tilde{q}^{2}\rangle$ respectively,
and $4\mathcal{P} = 4\mathcal{P}_U' + 4\mathcal{\tilde{P}}_U$.  The
term $4\mathcal{P}_{\tilde{u}}' \equiv -\langle\delta u'_i
({u'_j}^++{u'_j}^-) \frac{\partial \delta \tilde{u}_i}{\partial x_j}
\rangle - \langle 2\delta {u'_i} \delta {u'_j} \frac{\partial \delta
  \tilde{u}_i}{\partial r_j} \rangle$ appears with opposite signs in
\cref{eq:tripleKHM_stoch} and \cref{eq:tripleKHM_phase} and is
therefore the production term which exchanges energy at given ${\bf
  x}$ and ${\bf r}$ between the stochastic and the coherent
fluctuating motions. The spatial transport terms
$4\mathcal{T}_{\tilde{u}}'$ and $4\mathcal{T}_{u'}'$ are the first and
second terms in the third line of \cref{eq:tripleKHM_stoch} and the
stochastic pressure-stochastic velocity term $4\mathcal{T}_{p'}'$ is
the third term on this line. Similarly, the transport terms
$4\mathcal{\tilde{T}}_{\tilde{u}}$ and
$4\mathcal{\tilde{T}}_{\mathcal{P}_{\tilde{u}}}$ are the first and
second terms in the third line of \cref{eq:tripleKHM_phase} and the
coherent pressure-coherent velocity term
$4\mathcal{\tilde{T}}_{\tilde{p}}$ is the third term on this line. The
remaining terms are the diffusion terms $4\mathcal{\tilde{D}}_x$,
$4\mathcal{\tilde{D}}_r$, $4\mathcal{D}_x'$ and $4\mathcal{D}_r'$ and
the dissipation terms $4\tilde{\varepsilon}_r$ and $4\varepsilon'_r$
which are defined exactly as the diffusion and dissipation terms in
the KHMH \cref{eq:KHMH,eq:KHMHsimple} but for the coherent and
stochastic velocity fields, respectively, rather than for the total
fluctuating velocity field.

Adding \cref{eq:tripleKHMsimple_stoch} with
\cref{eq:tripleKHMsimple_phase} results in the KHMH equation by
combining terms with tilde and primes together (e.g. $\mathcal{A}=
\mathcal{A}' + \mathcal{\tilde{A}}$, $\Pi_U = \Pi_U' + \tilde{\Pi}_U$,
etc) but also by noticing that
\begin{equation}\label{eq:total_Pi_tmp}
\Pi=\Pi'+\Pi_{\tilde{u}}'+ \tilde{\Pi}_{\tilde{u}}+\tilde{\Pi}_{\mathcal{P}_{\tilde{u}}}
\end{equation}
and
\begin{equation}\label{eq:total_Tu}
\mathcal{T}_{u}=\mathcal{T}_{u'}'+\mathcal{T}_{\tilde{u}}'+
\tilde{\mathcal{T}}_{\tilde{u}}+\tilde{\mathcal{T}}_{\mathcal{P}_{\tilde{u}}} 
\end{equation}
which are the non-linear inter-scale and inter-space transfer terms.

The terms $\Pi'$, $\Pi_{\tilde{u}}'$ and $\tilde{\Pi}_{\tilde{u}}$ can
be interpreted as inter-scale transfer terms of either $\delta {q'}^2$
or $\delta \tilde{q}^2$. $4\Pi'\equiv \frac{\partial}{\partial
  r_i}\langle\delta u'_i \delta {q'}^2\rangle $ represents the
inter-scale transfer of energy associated with the stochastic motions
by the stochastic motions (i.e. inter-scale transfer of $\delta
{q'}^2$ by $\delta {\bf u}'$). Similarly, $4\Pi'_{\tilde{u}} \equiv
\frac{\partial}{\partial r_i}\langle\delta \tilde{u}_i \delta
     {q'}^2\rangle $ represents the inter-scale transfer of the energy
     associated with the stochastic motions by the coherent motions
     (i.e. inter-scale transfer of $\delta {q'}^2$ by $\delta
      \tilde{\mathbf{u}}$), and $4\tilde{\Pi}_{\tilde{u}}\equiv
     \frac{\partial}{\partial r_i}\langle\delta \tilde{u}_i \delta
     \tilde{q}^2\rangle$ represents the inter-scale transfer of the
     energy associated with the coherent motions by the coherent
     motions (i.e. inter-scale transfer of $\delta \tilde{q}^2$ by
     $\delta \tilde{u}_i$). The term
     $4\tilde{\Pi}_{\mathcal{P}_{\tilde{u}}}$ can be written as the
     difference between two inter-scale transfer terms: the
     inter-scale transfer by the stochastic velocity field of the
     total fluctuating energy and $4\Pi'$, i.e.
     $4\tilde{\Pi}_{\mathcal{P}_{\tilde{u}}} = 4\Pi_{u'} - 4\Pi'$
     where $4\Pi_{u'} \equiv \frac{\partial}{\partial r_i}
     \langle\delta u'_i |\delta \mathbf{u}' + \delta
     \mathbf{\tilde{u}}|^2 \rangle$. Hence, combining
     $\tilde{\Pi}_{\mathcal{P}_{\tilde{u}}}$ with $\Pi'$ results in
     the inter-scale transfer of the total fluctuating energy by the
     stochastic motions (i.e. inter-scale transfer of $\delta q^2$ by
     $\delta \mathbf{u}'$) so that \cref{eq:total_Pi_tmp} can be
     written as
\begin{equation}\label{eq:total_Pi}
\Pi=\Pi_{u'}+\Pi_{\tilde{u}}'+\tilde{\Pi}_{\tilde{u}}
\end{equation}
This proves to be an important equation in \cref{subsec:tripledecomp_constPi}.

The terms $\mathcal{T}_{u'}'$, $\mathcal{T}_{\tilde{u}}'$,
$\mathcal{\tilde{T}}_{\tilde{u}}$ represent turbulent transport in
physical space. Specifically, $4\mathcal{T}_{u'}' \equiv
-\langle\frac{{u'_i}^++{u'_i}^-}{2}\frac{\partial \delta
  {q'}^2}{\partial x_i}\rangle$ represents inter-space transport of
stochastic turbulent energy by stochastic fluctuations,
$4\mathcal{T}_{\tilde{u}}' \equiv
-\langle\frac{{\tilde{u}_i}^++{\tilde{u}_i}^-}{2}\frac{\partial \delta
  {q'}^2}{\partial x_i}\rangle$, represents inter-space transport of
stochastic turbulent energy by coherent fluctuations, and
$\mathcal{\tilde{T}}_{\tilde{u}} \equiv
-\langle\frac{{\tilde{u}_i}^++{\tilde{u}_i}^-}{2}\frac{\partial \delta
  {\tilde{q}}^2}{\partial x_i}\rangle$ represents inter-space
transport of coherent fluctuating energy by coherent fluctuations.
The term $-4\mathcal{\tilde{T}}_{\mathcal{P}_{\tilde{u}}}$ is the
difference between $4\mathcal{T}_{u'}'$ and the spatial transport of
the total fluctuating energy by the two-point-average stochastic
velocity, i.e. $4\mathcal{\tilde{T}}_{\mathcal{P}_{\tilde{u}}} =
4\mathcal{T}_{u'} - 4\mathcal{T}_{u'}'$ were $4\mathcal{T}_{u'} \equiv
\langle\frac{{u'_i}^++{u'_i}^-}{2}\frac{\partial \vert \delta
  \mathbf{u}' + \delta \mathbf{\tilde{u}}\vert^2}{\partial
  x_i}\rangle$.  This allows rewriting \cref{eq:total_Tu} as follows:
\begin{equation}\label{eq:total_Tu2}
\mathcal{T}_{u}=\mathcal{T}_{u'}+\mathcal{T}_{\tilde{u}}'+
\tilde{\mathcal{T}}_{\tilde{u}} .
\end{equation}

In the following section we compare the signs and magnitudes of the
orientation-averaged terms in \cref{eq:tripleKHMsimple_stoch} and
\cref{eq:tripleKHMsimple_phase} in the
near field turbulent planar wake.

\section{Orientation-averaged scale-by-scale energy budgets in the near wake of a square prism}\label{sec:orientation-averaged}

Each term, $Q$, in \cref{eq:tripleKHMsimple_stoch} and
\cref{eq:tripleKHMsimple_phase} is an average in time and span-wise
direction and is therefore a function of planar coordinates $(x_{1},
x_{2})$ and two-point separation vector ${\bf r}$,
i.e. $Q=Q(x_{1},x_{2}, {\bf r})$. We set $r_{3}=0$ and define the
orientation-averaged quantity $Q^{a}$ by integrating $Q$ over the
angle $\theta$ defined by $r_{1} = r\cos\theta$, $r_{2} = r\sin\theta$
which also defines the radius (and length-scale) $r$: $Q^{a} (x_{1},
x_{2}, r) \equiv {1\over 2\pi}\int_{0}^{2\pi} Q d\theta$. Such
scale-space orientation-averaging has already been used by
\citet{AlvesPortela2017} and \citet{Gomes-Fernandes2015} to study the
terms in the KHMH \cref{eq:KHMHsimple}. We verified that the KHMH
equation \cref{eq:KHMH} is sufficiently well balanced numerically, as
the difference between its left hand and right hand sides is two
orders of magnitude smaller than $\varepsilon_{r}$ for all $r$
investigated here, and even smaller than that when the two sides are
orientation-averaged in scale-space plane $r_{3}=0$. We also checked
that every term in \cref{eq:KHMH} is indeed equal to the sum of its
two corresponding terms in equations \cref{eq:tripleKHM_stoch} and
\cref{eq:tripleKHM_phase}, for example $\mathcal{A}= \mathcal{A}' +
\mathcal{\tilde{A}}$, $\Pi_U = \Pi_U' + \tilde{\Pi}_U$, etc.

\begin{figure}
\centering
\begin{subfigure}[t]{\textwidth}
\centering
\includegraphics{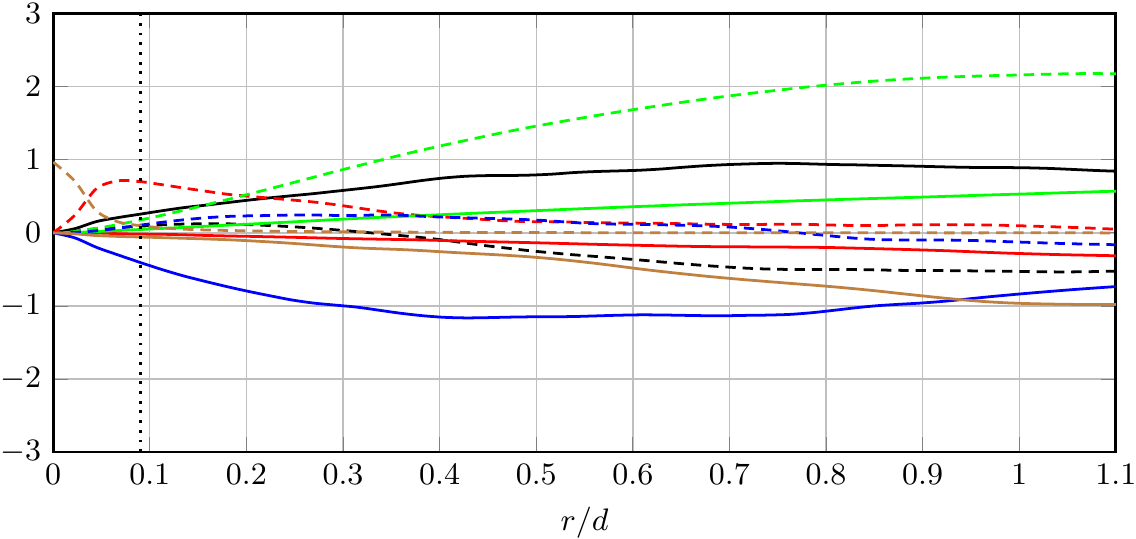}
\end{subfigure}
\begin{subfigure}[t]{\textwidth}
\centering
\includegraphics{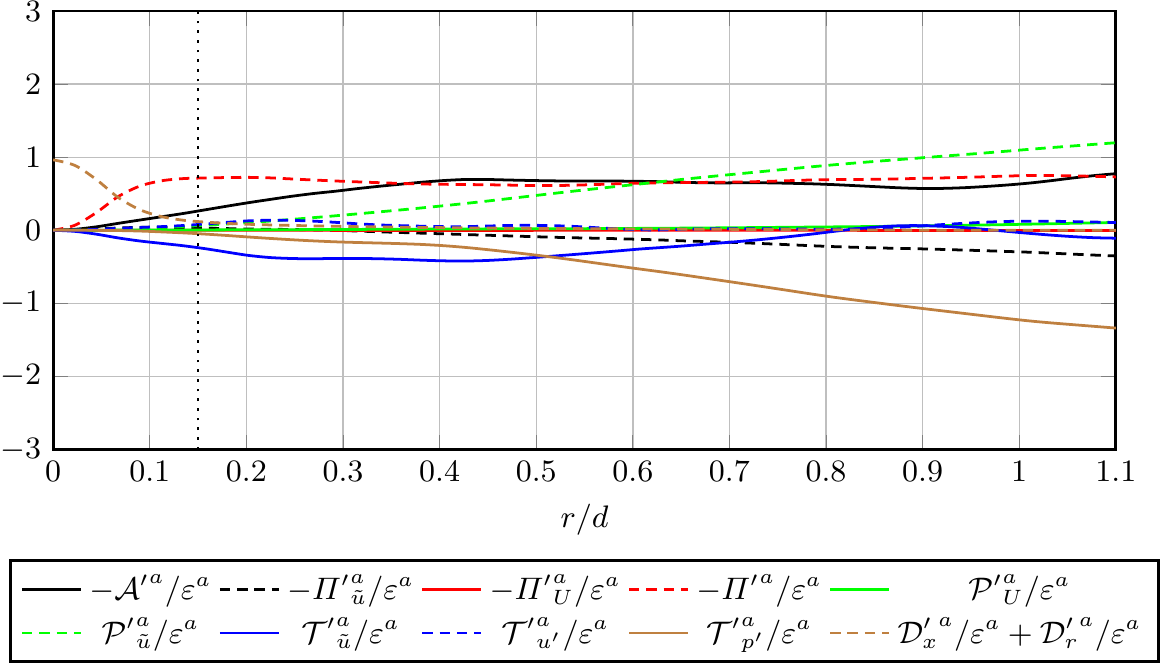}
\end{subfigure}
\caption{Orientation-averaged terms of \cref{eq:tripleKHMsimple_stoch}
  (equivalently \cref{eq:tripleKHM_stoch}) normalised by
  $\varepsilon^a$ at $x_1/d=2$ (top) and $x_1/d=8$ (bottom) on the
  geometric centreline. The vertical dotted line gives the position of
  $r=\lambda$.}\label{fig:stochKHMH_ravg_x28}
\end{figure}

In \cref{fig:stochKHMH_ravg_x28} we plot all the orientation-averaged
terms in \cref{eq:tripleKHMsimple_stoch} versus $r/d$ in the range
$0\le r/d \le 1.1$ at two centreline positions, $(x_{1},
x_{2})=(2d,0)$ and $(8d,0)$. These terms are plotted normalised by
$\varepsilon_{r}^a$ which, for $r$ not much larger than $d$, is
approximately equal to $\varepsilon_{r}'^a$
\citep[see][]{AlvesPortela2018a} and to the one-point dissipation rate
$\varepsilon$ in the region of the centreline that we study. The range
$0\le r\le 1.1d$ has also been chosen because the average distance
between consecutively shed coherent vortices is comparable to $3d$.

The first observation to make in \cref{fig:stochKHMH_ravg_x28} is that
the near wake region is so inhomogeneous that most of the terms in the
scale-by-scale energy budget \cref{eq:tripleKHMsimple_stoch} are
active. The terms dominating the range $0.4\le r/d \le 1.1$ at
$x_{1}/d=2$ are $-\mathcal{A}'^{a}$ ($\mathcal{A}'\equiv
\frac{U_i^++U_i^-}{2} \frac{\partial}{\partial x_i}\langle \delta
     {q'}^2 \rangle$) and $\mathcal{P}_{\tilde{u}}'^{a}$ ($
     \mathcal{P}_{\tilde{u}}'\equiv - \langle\delta u'_i
     ({u'_j}^++{u'_j}^-) \frac{\partial \delta \tilde{u}_i}{\partial
       x_j} \rangle-2\langle \delta {u'_i} \delta {u'_j}
     \frac{\partial \delta \tilde{u}_i}{\partial r_j} \rangle$) which
     are both positive, and $\mathcal{T}_{\tilde{u}}'^{a}$
     ($\mathcal{T}_{\tilde{u}}'\equiv -
     \langle\frac{\tilde{u}_i^++\tilde{u}_i^-}{2}\frac{\partial \delta
       {q'}^2}{\partial x_i}\rangle
     -\langle\frac{{u'_i}^++{u'_i}^-}{2}\frac{\partial \delta
       {q'}^2}{\partial x_i}\rangle$) and $\mathcal{T}_{p'}'^{a}$
     ($\mathcal{T}_{p'}' \equiv - 2\langle \delta u'_i \delta
     \frac{\partial p'}{\partial x_i}\rangle$) which are both negative
     (positive/negative terms correspond to a gain/loss in the
     budget). These terms are closely followed by the production of
     stochastic turbulent fluctuations by mean flow gradients,
     $\mathcal{P}_U'^{a}$ ($\mathcal{P}_U'\equiv -\langle\delta u'_i
     ({u'_j}^++{u'_j}^-) \rangle \frac{\partial \delta U_i}{\partial
       x_j} -2 \langle \delta u'_i \delta u'_j \rangle \frac{\partial
       \delta U_i}{\partial r_j}$), which is positive, and by the
     inter-scale transfer of stochastic fluctuating energy by coherent
     motions, plotted with a minus sign as $-\Pi_{\tilde{u}}'^{a}$
     ($\Pi_{\tilde{u}}' \equiv \frac{\partial}{\partial
       r_i}\langle\delta \tilde{u}_i \delta {q'}^2\rangle$), which is
     negative. The term which is in fact the largest in this
     scale-range at $x_{1}/d=2$ is $\mathcal{P}_{\tilde{u}}'^{a}$, the
     rate of energy transfer between the coherent and stochastic
     fluctuating motions. This term being positive for all values of
     $r$ in \cref{fig:stochKHMH_ravg_x28}, the coherent motions feed
     energy to the stochastic ones at all these scales. At the same
     time, the coherent fluctuations are responsible for removing
     energy from the stochastic ones by spatial transport;
     $\mathcal{T}_{\tilde{u}}'^{a}$ is negative and dominant at all
     scales $r$ too. Recall that these scale-dependent energy
     exchanges happen at $x_{1}/d =2$ on the centreline where the
     energy spectra have a broad well-defined power law range with
     exponent close to $-5/3$ (see figure \cref{fig:Euv_phasestoch})
     as already shown by \citet{AlvesPortela2017}.

Note that the orientation-averaged non-linear inter-scale transfer
rate, plotted with a minus sign as $-\Pi'^{a}$ ($\Pi' \equiv
\frac{\partial}{\partial r_i}\langle\delta u'_i \delta
     {q'}^2\rangle$), is not too significant in the range $0.4\le r/d
     \le 1.1$ at $(x_{1}, x_{2})=(2d, 0)$. However it is one of the
     four dominant terms in the range $\lambda/d \le r/d \le 0.4$ at
     this location. These four dominant terms are $-\mathcal{A}'^{a}$,
     $-\Pi'^{a}$, $\mathcal{P}_{\tilde{u}}'^{a}$ and
     $\mathcal{T}_{u'}'^{a}$, and $\lambda$ is the Taylor microscale
     defined as $\lambda^{2} \equiv 2\langle u_{3}^{2}\rangle /
     \langle(\frac{\partial}{\partial x_{3}} u_{3})^{2}\rangle$. At
     $(x_{1}, x_{2})=(2d, 0)$ $\lambda$ is 0.09d
and at $(x_{1}, x_{2})=(8d, 0)$ $\lambda$ is
$0.15d$. The diffusion terms $\mathcal{D}_x'^{a}$ and
$\mathcal{D}_r'^{a}$ effectively vanish at length-scales $r$ larger
than $\lambda$, and they equal $\varepsilon'^{a}$ at $r=0$, as
expected \citep[see][]{Valente2013}.

It is worth stressing that, at $(x_{1}, x_{2})=(2d, 0)$, the
orientation-averaged non-linear inter-scale transfer rate $\Pi'^{a}$
is mainly balanced by the advection term $-\mathcal{A}'^{a}$ and
coherent motion production and transport processes,
i.e. $\mathcal{P}_{\tilde{u}}'^{a}$ and $\mathcal{T}_{\tilde{u}}'^{a}$, in
the range $\lambda \le r \le 0.4d$. Even though energy spectra have
well-defined power law ranges with exponents close to $-5/3$ at
$(x_{1}, x_{2})=(2d, 0)$, $\Pi'^{a}$ is not constant with length-scale
$r$.

Further downstream, at $(x_{1}, x_{2})=(8d, 0)$, the
orientation-averaged non-linear inter-scale transfer rate $\Pi'^{a}$
is mainly balanced by the advection term $-\mathcal{A}'^{a}$ and
coherent motion transport , i.e. $\mathcal{T}_{\tilde{u}}'^{a}$, in
the range $\lambda \le r \le 0.3d$. All the other orientation-averaged
terms in \cref{eq:tripleKHMsimple_stoch} are less significant in this
scale-range and at this position. The orientation-averaged impact of
the coherent motions on the scale-by-scale budget
\cref{eq:tripleKHMsimple_stoch} gradually diminishes with increasing
distance from the square prism. In the range $0.3d \le r \le 1.1d$ at
$(x_{1}, x_{2})=(8d, 0)$, the dominant terms are now
$-\mathcal{A}'^{a}$, $\mathcal{P}_{\tilde{u}}'^{a}$ and $-\Pi'^{a}$
which are all still positive, and $\mathcal{T}_{p'}'^{a}$ which is
still negative. The term $\mathcal{T}_{\tilde{u}}'^{a}$ has greatly
reduced in relative importance from $(x_{1}, x_{2})=(2d, 0)$ to
$(x_{1}, x_{2})=(8d, 0)$, but the presence of the pressure-velocity
term $\mathcal{T}_{p'}'^{a}$ has remained significant and about the
same, if not even grown a little. Perhaps most striking of all is the
fact that $-\Pi'^{a}$ has grown to become closer to an approximate
constant fraction of $\varepsilon_{r}^a$ in the range $\lambda\le r
\le 1.1d$ at $(x_{1}, x_{2})=(8d, 0)$ which is downstream of the point
where the near $-5/3$ power law spectra appeared.

\begin{figure}
    \centering
    \begin{subfigure}[b]{\textwidth}
        \centering
        \includegraphics{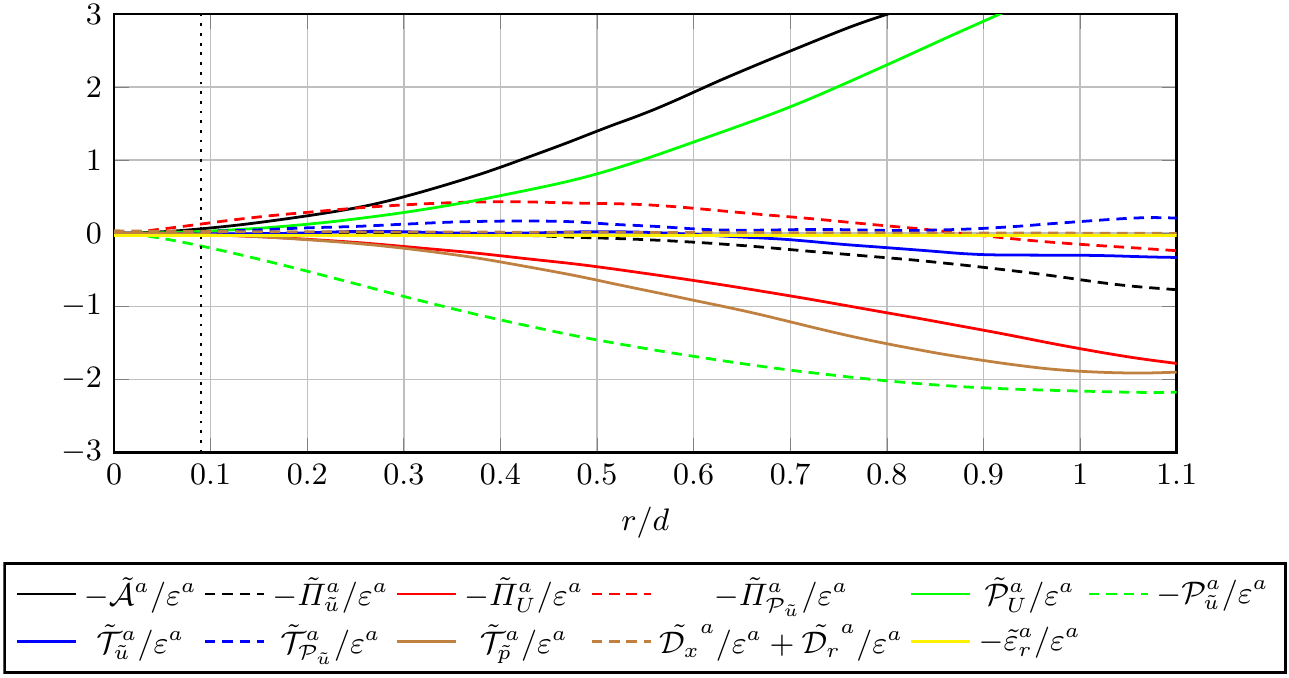}
    \end{subfigure}
    \centering
    \begin{subfigure}[b]{\textwidth}
        \centering
        \includegraphics{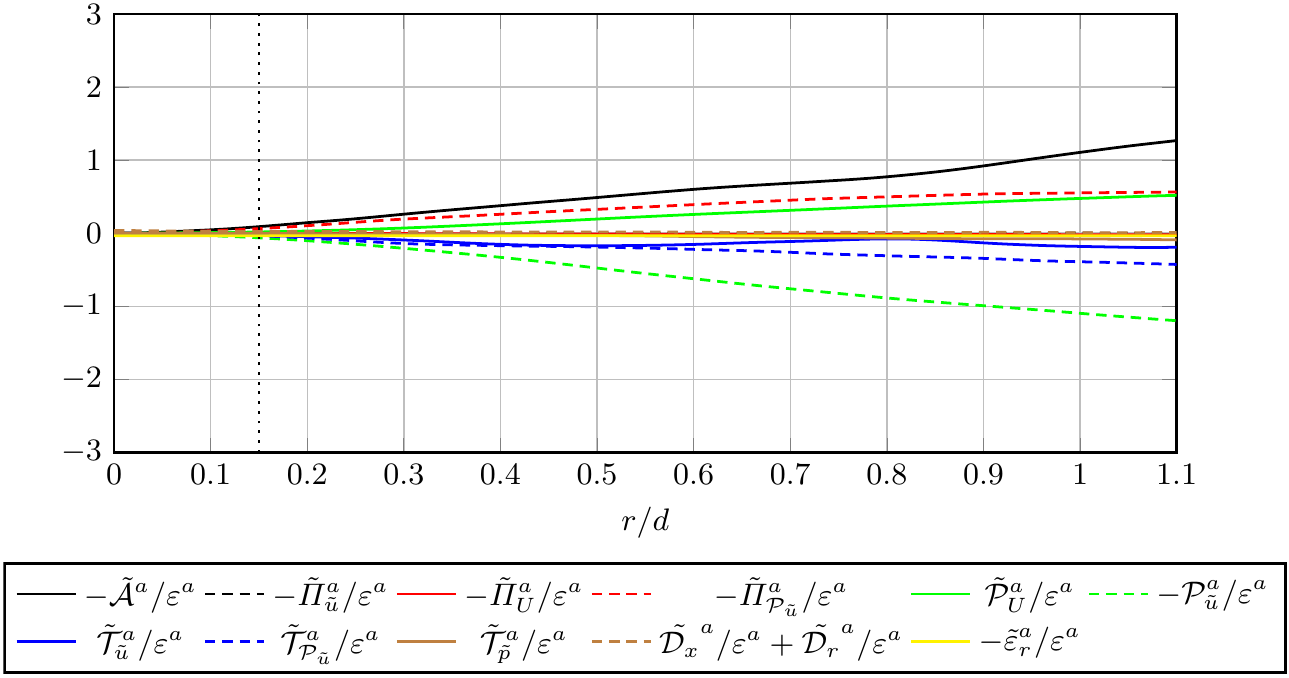}
    \end{subfigure}
     \caption{Orientation-averaged terms of
       \cref{eq:tripleKHMsimple_phase} (equivalently
       \cref{eq:tripleKHM_phase}) normalised by $\varepsilon_r^a$ at
       $x_1/d=2$ (top) and $x_{1}/d=8$ (bottom) on the geometric
       centreline. The vertical dotted line gives the position of
       $r=\lambda$.}\label{fig:phaseKHMH_ravg_x2}
\end{figure}

\begin{figure}
    \centering
    \begin{subfigure}[b]{\textwidth}
        \centering
        \includegraphics{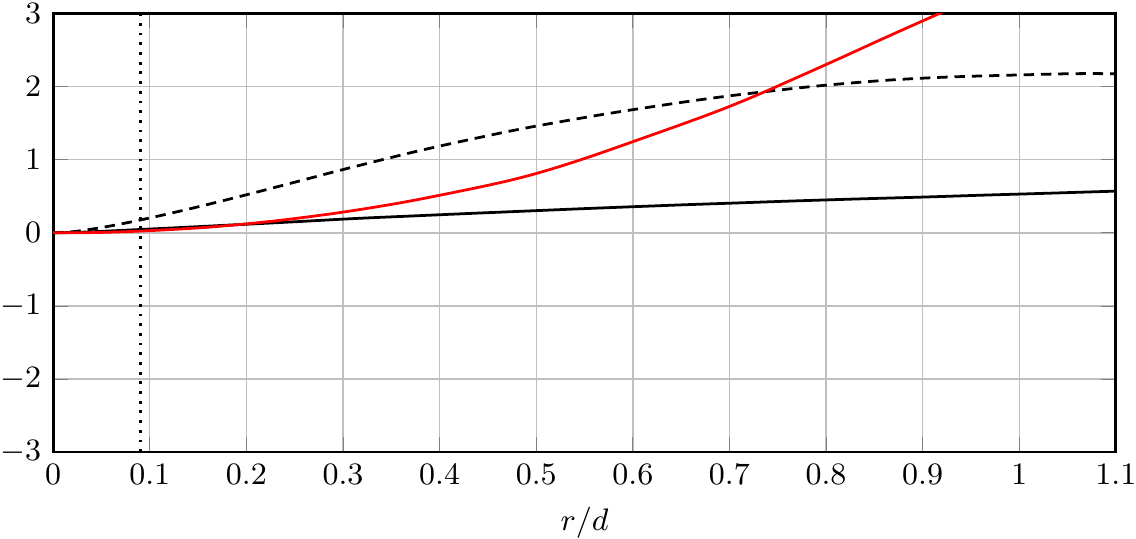}
    \end{subfigure}
    \centering
    \begin{subfigure}[b]{\textwidth}
        \centering
        \includegraphics{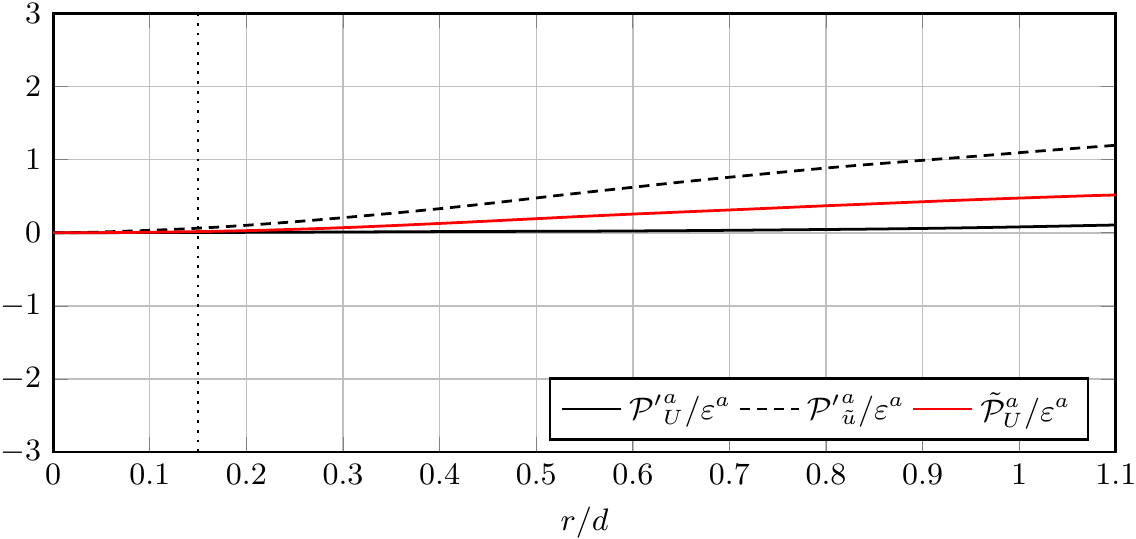}
    \end{subfigure}
     \caption{Orientation-averaged production terms
         normalised by $\varepsilon_r^a$ at $x_1/d=2$ (top) and
         $x_{1}/d=8$ (bottom) on the geometric centreline. The
         vertical dotted line gives the position of
         $r=\lambda$.}\label{fig:PRODUCTIONS}
\end{figure}

Production of stochastic fluctuation energy by mean flow gradients,
namely $\mathcal{P}_U'^{a}$, is a minor contributor to the
scale-by-scale stochastic fluctuation balance
\cref{eq:tripleKHMsimple_stoch} at $(x_{1}, x_{2})=(2d, 0)$ and
effectively inexistent at $(x_{1}, x_{2})=(8d, 0)$ (see
\cref{fig:stochKHMH_ravg_x28}). However, \cref{fig:phaseKHMH_ravg_x2}
shows that production of coherent scale-by-scale energy by mean flow
gradients, specifically $\mathcal{\tilde{P}}_U^a$
($\mathcal{\tilde{P}}_U \equiv -\langle\delta \tilde{u}_i
(\tilde{u}_j^++\tilde{u}_j^-)\rangle \frac{\partial \delta
  U_i}{\partial x_j} -2\langle \delta \tilde{u}_i \delta \tilde{u}_j
\rangle \frac{\partial \delta U_i}{\partial r_j}$), is an important
source of scale-by-scale energy in the coherent fluctuations balance
\cref{eq:tripleKHMsimple_phase} at both positions $(x_{1}, x_{2})=(2d,
0)$ and $(8d, 0)$. A clear picture emerges whereby, in an
orientation-averaged sense, the mean flow gradients do not
significantly feed the stochastic fluctuations directly but do feed
the coherent motions which, in turn, feed the stochastic fluctuations
via $\mathcal{P}_{\tilde{u}}'^{a}$. Indeed, the term
$\mathcal{P}_{\tilde{u}}'^{a}$ appears as a dominant term in the
orientation-averaged versions of both budgets
\cref{eq:tripleKHMsimple_stoch} and \cref{eq:tripleKHMsimple_phase}
(see \cref{fig:PRODUCTIONS}, and also \cref{fig:stochKHMH_ravg_x28}
and \cref{fig:phaseKHMH_ravg_x2}) but with opposite signs. This holds
over a wide range of scales as small as $\lambda$ for the transfer of
energy from the coherent to the stochastic fluctuations at $(x_{1},
x_{2})=(2d, 0)$ and as small as about $2\lambda$ or less for the
production by mean flow gradients at $(x_{1}, x_{2})=(2d, 0)$ and for
both $\mathcal{P}_{\tilde{u}}'^{a}$ and $\mathcal{\tilde{P}}_U^a$ at
$(x_{1}, x_{2})=(8d, 0)$ (see figure \cref{fig:PRODUCTIONS}).

\begin{figure}
    \centering
    \begin{subfigure}[b]{\textwidth}
        \centering
        \includegraphics{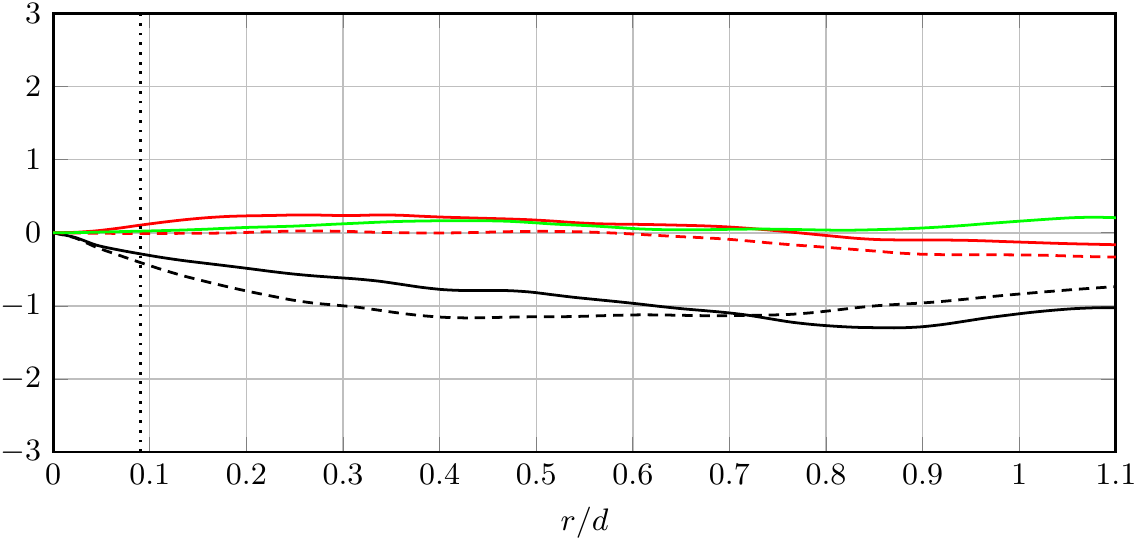}
    \end{subfigure}
    \centering
    \begin{subfigure}[b]{\textwidth}
        \centering
        \includegraphics{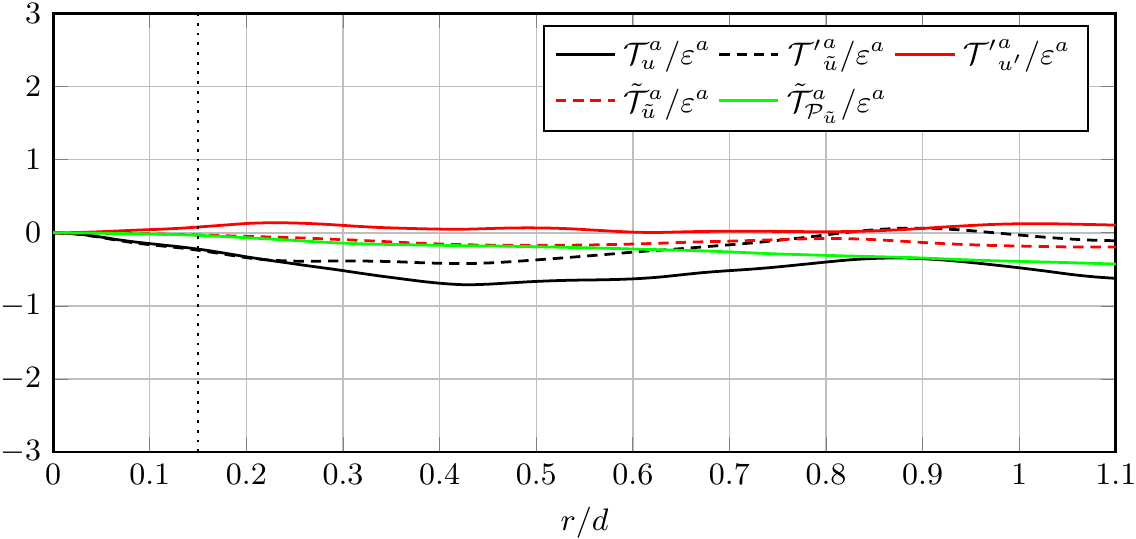}
    \end{subfigure}
     \caption{Orientation-averaged spatial transport terms normalised
       by $\varepsilon_r^a$ at $x_1/d=2$ (top) and $x_{1}/d=8$
       (bottom) on the geometric centreline. The vertical dotted line
       gives the position of $r=\lambda$.}\label{fig:TRANSPORTS}
\end{figure}

The terms in \cref{eq:tripleKHMsimple_phase} mostly decay with
stream-wise distance from the prism along the centreline (see
\cref{fig:phaseKHMH_ravg_x2}), but they remain overall comparable to
the terms in \cref{eq:tripleKHMsimple_stoch} at the two positions
$(x_{1}, x_{2})$ examined here, particularly at length-scales $r\ge
0.2d$ or $0.3d$.  Looking at \cref{eq:total_Tu} and
\cref{fig:TRANSPORTS}
we can see that the orientation-averaged turbulent transport of
$\delta q^2$ in physical space ($\mathcal{T}_u^{a} \equiv
-{\frac{\partial \langle \frac{u_i^++u_i^-}{2} \delta
    q^2\rangle}{\partial x_i}}$) is dominated by the
orientation-averaged transport of stochastic fluctuations by coherent
flow, i.e. $\mathcal{T}'^{a}_{\tilde{u}}$, at $(x_{1}, x_{2})=(2d, 0)$
over all plotted length-scales $r$ and at $(x_{1}, x_{2})=(8d, 0)$ up
to $r/d \approx 0.5$. Indeed, the fluid between alternate coherent
vortices (of opposite circulation) has large cross-stream velocities
which dominate turbulent transport in space.
$\mathcal{T}’^{a}_{\tilde{u}}$ is negative because turbulent eddies
smaller than the separation between these large-scale coherent
vortices are transported away from the centreline. We expect this
dominance of coherent flow transport to subside with downstream
distance as the large coherent structures weaken.

\begin{figure}
    \centering
    \begin{subfigure}[t]{0.45\textwidth}
        \centering
        \includegraphics{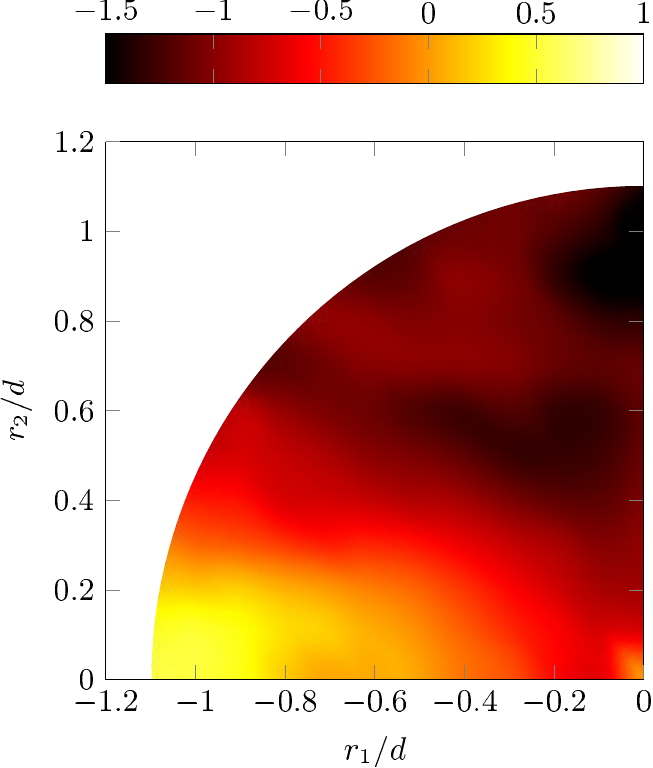}
    \captionsetup{skip=-1.2\textwidth}
\caption{\hspace*{0.8\textwidth}}\label{fig:Pisss_x8_zoom}
    \end{subfigure}%
    \hspace*{0.05\textwidth} 
    \begin{subfigure}[t]{0.45\textwidth}
        \centering
        \includegraphics{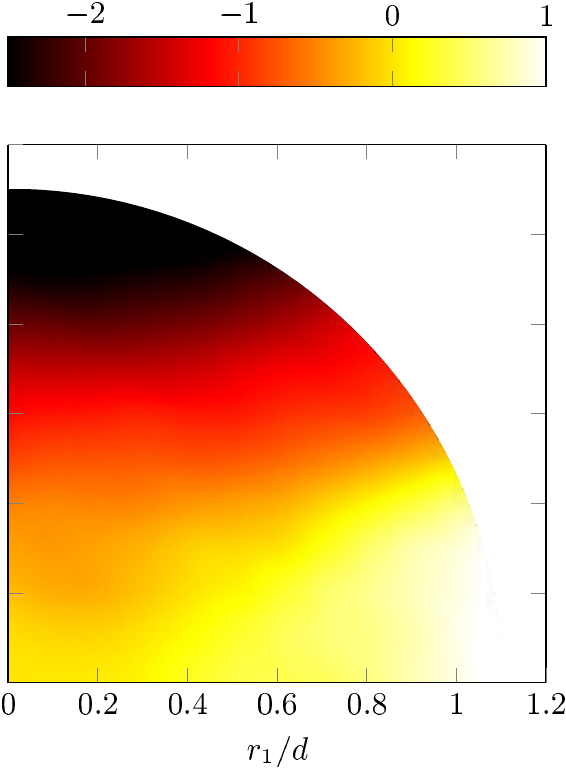}
    \captionsetup{skip=-1.2\textwidth}
\caption{\hspace*{-\textwidth}}\label{fig:Tpss_x8_zoom}
    \end{subfigure}
    \caption{Distribution of $\Pi'$ and $\mathcal{T}_{p'}'$ normalised
      by $\varepsilon_r$ in scale space on the geometrical centreline
      at $x_1/d=8$ in (a) and (b),
      respectively.}\label{fig:Pisss_Tpss_x8_zoom}
\end{figure}

The results reported in this section concern orientation-averaged
terms of equations (\ref{eq:tripleKHMsimple_phase}) and
(\ref{eq:tripleKHMsimple_stoch}). The picture is of course more
complex if these orientation averages are lifted. For example, the
orientation-averaged fully stochastic non-linear inter-scale transfer
rate $\Pi'^{a}$ is negative at all length-scales $r$ sampled here, yet
$\Pi'$ can be either negative or positive in the $(r_{1}, r_{2})$
plane, depending on orientation (see
\cref{fig:Pisss_Tpss_x8_zoom}). Similarly, the orientation-averaged
fully stochastic pressure-velocity term $\mathcal{T}_{p'}'^{a}$ is
also negative at all the length-scales $r$ that we sampled, yet
$\mathcal{T}_{p'}'$ can also be either negative or positive in the
$(r_{1}, r_{2})$ plane depending on orientation, as shown in
\cref{fig:Pisss_Tpss_x8_zoom}. The study of the distribution in the
$(r_{1}, r_{2})$ plane of the various terms in equations
(\ref{eq:tripleKHMsimple_phase}) and (\ref{eq:tripleKHMsimple_stoch})
is beyond this paper's scope, but it is worth noting the correlation
that seems to exist between $\Pi'$ and $\mathcal{T}_{p'}'$:
\cref{fig:Pisss_Tpss_x8_zoom} shows a significant tendency for these
two terms to be positive or negative together. A correlation between
fluctuations of the non-linear inter-scale transfer rate and the
pressure-velocity term has also been observed in DNS of periodic
turbulence by \citet{Yasuda2018} where it is discussed in
more detail.

\section{Effects of the Coherent motion and inhomogeneity on the Inter-scale Energy Transfer}\label{subsec:tripledecomp_constPi}

\citet{AlvesPortela2017} showed how the average non-linear inter-scale
transfer rate of $\delta {q}^2$ is roughly constant when the
orientations of $\mathbf{r}$ are averaged out in the $r_3=0$ plane,
despite this transfer rate's distribution being far from uniform in
this plane. This was in fact observed in spite of the severe
inhomogeneities and anisotropies evidenced in the previous section by
the various non-zero terms in the KHMH equations
(\ref{eq:tripleKHMsimple_phase}) and (\ref{eq:tripleKHMsimple_stoch}),
and even at $x_1/d=2$ (albeit for a small range of separations) where
the coherent motions contribute a large portion of the total
fluctuating kinetic energy (recall
\cref{fig:tke_phase_centreline}). In this section we start by
determining how this constancy of $\Pi^a$ observed in
\citet{AlvesPortela2017} and in \citet{Gomes-Fernandes2015} depends on
contributions arising from the coherent and stochastic motions
individually (\cref{subsec:constPI_triple}), but also on statistical inhomogeneity
(\cref{subsec:inhomog_scale_transfer}). 
We close the section by checking the signs of inter-scale fluxes in
\cref{subsubsec:fluxes}.

\subsection{Constant Non-linear Inter-scale Transfer as a Combined Effect}\label{subsec:constPI_triple}

\begin{figure}
    \centering
    \begin{subfigure}[b]{\textwidth}
        \centering
        \includegraphics{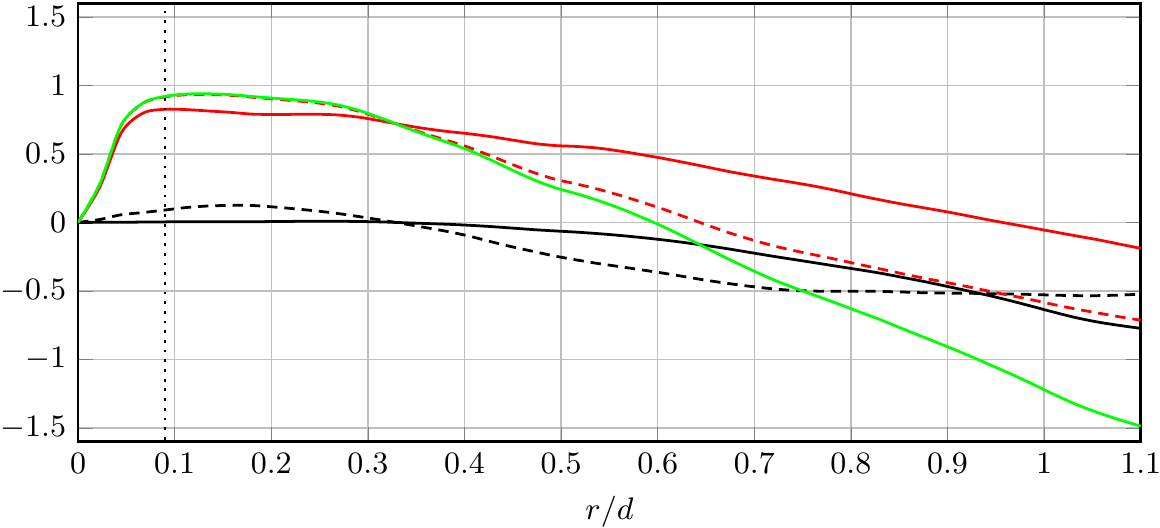}
    \end{subfigure}%
\\
    \begin{subfigure}[b]{\textwidth}
        \centering
        \includegraphics{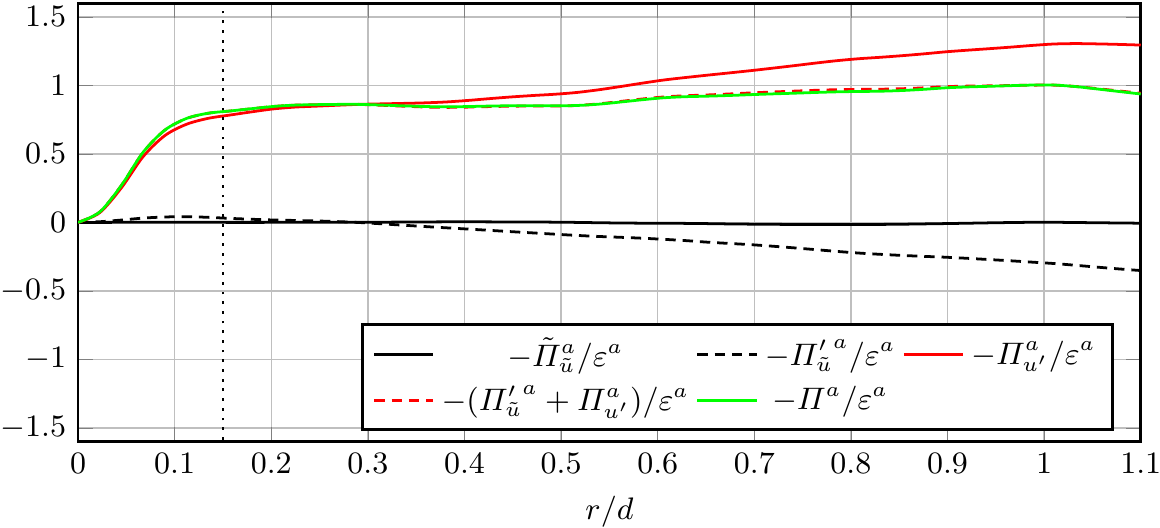}
    \end{subfigure}
    \caption{Orientation averaged non-linear inter-scale transfer
      terms from \cref{eq:tripleKHM_phase,eq:tripleKHM_stoch}
      normalised by $\varepsilon_r^a$ at $x_1/d=2$ (top) and $x_1/d=8$
      (bottom). The vertical dotted line gives the position of
      $r=\lambda$.}\label{fig:Pi_stochphase_ravg_x28}
\end{figure}

As mentioned in the previous section, the orientation-averaged
inter-scale transfer rate of stochastic fluctuating energy by
stochastic motions, $\Pi'^{a}$, is not independent of length-scale $r$
at $x_1/d=2$ on the centreline. However,
\cref{fig:Pi_stochphase_ravg_x28} shows that $\Pi_{u'}^{a}$, the
orientation-averaged inter-scale transfer rate of total fluctuating
energy by the stochastic motions is close to being constant with $r$
in the range $\lambda < r < 0.3d$ at this point ($x_1/d=2$,
$x_{2}/d=0$). Furthermore, this approximate constant is closer to
$-\varepsilon_{r}^{a}$ if $\Pi_{\tilde{u}}'^{a}$ is taken into
account, i.e. at $(x_{1}, x_{2})=(2d,0)$, $\Pi_{u'}^{a} +
\Pi_{\tilde{u}}'^{a}$ is also approximately constant in the range
$\lambda < r < 0.3d$ and closer to $-\varepsilon_{r}^{a}$ than
$\Pi_{u'}^{a}$. In fact, at this location,
$\tilde{\Pi}_{\tilde{u}}^a\approx 0$ and \cref{eq:total_Pi} reduces to
\begin{equation}\label{eq:total_Pi_notilde}
\Pi^a\approx\Pi_{u'}^a+{\Pi_{\tilde{u}}'}^a
\end{equation}
in this range where $\Pi^a$ is approximately constant and close to
$-\varepsilon_{r}^a$ (which is, in fact, very closely equal to
$-\varepsilon$ in this range). This \cref{eq:total_Pi_notilde} also
holds further downstream on the centreline, but over a longer range of
scales, e.g.  $\lambda <r< d$ at $(x_{1}, x_{2})=(8d,0)$ (see
\cref{fig:Pi_stochphase_ravg_x28}).

The fact that a scale-range exists where $\Pi^a/\varepsilon^a_r$ is
approximately constant and relatively close to $-1$ would not have been
possible without the presence of coherent structures at $(x_{1},
x_{2})=(2d,0)$. Whilst these coherent structures are non-dynamic in
this scale-range, in the sense that $\tilde{\Pi}_{\tilde{u}}^a\approx
0$, they contribute to this clearly non-Kolmogorov yet
Kolmogorov-sounding approximately constant value of
$\Pi^{a}/\epsilon_{r}^{a}$ close to $-1$ in two ways: predominantly
through $\Pi_{u'}^a$ for the constancy of $\Pi^a/\varepsilon^a_r$, and
through ${\Pi_{\tilde{u}}'}^a$, the inter-scale transfer rate of
stochastic energy by coherent fluctuations which improves the
proximity of $\Pi^a/\varepsilon^a_r$ to $-1$.

\begin{figure}
        \centering
        \includegraphics{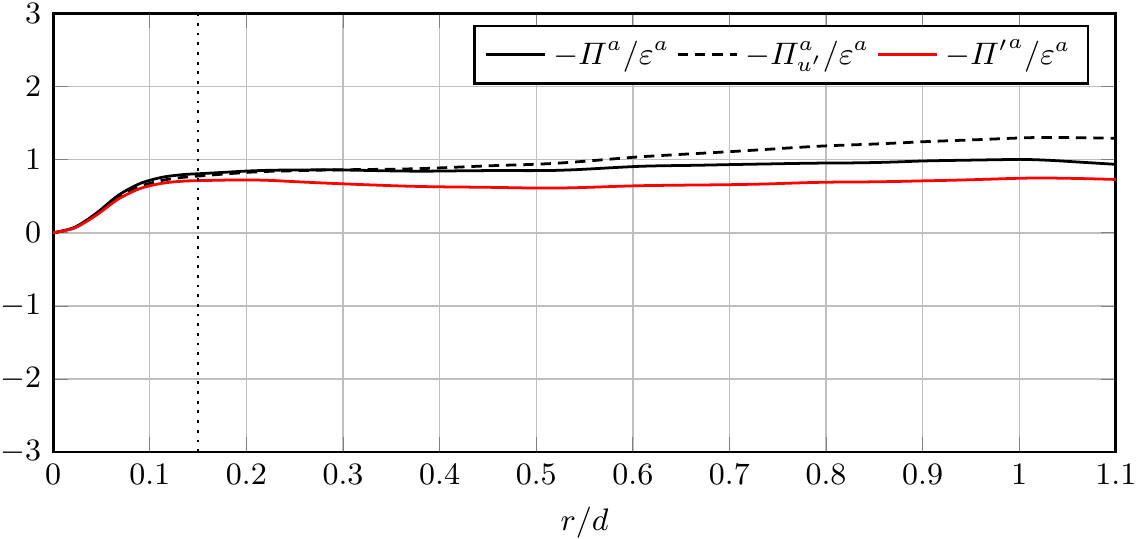}
     \caption{Orientation-averaged inter-scale transfer terms
       normalised by $\varepsilon_r^a$ at $x_{1}/d=8$ on the geometric
       centreline. The vertical dotted line gives the position of
       $r=\lambda$.}\label{fig:PI}
\end{figure}

Further downstream, at $(x_{1}, x_{2})=(8d,0)$,
$\Pi^a\approx\Pi_{u'}^a$ in the range $\lambda <r< 0.4d$. In this
range and at this position, the orientation-averaged inter-scale
transfer rate of stochastic energy by coherent fluctuations is zero,
and the near-constancy with scale $r$ of $\Pi^a$ is in fact, to a
significant extent, accountable to $\Pi'^a$, the orientation-averaged
inter-scale transfer rate of stochastic energy by stochastic
fluctuations (see \cref{fig:stochKHMH_ravg_x28}). But the coherent
structures also contribute significantly because $\Pi^a$ is slightly
but not insignificantly different from $\Pi'^a$, in such a way that
$\Pi^a\approx\Pi_{u'}^a$ is markedly closer to a constant than
$\Pi'^a$ in this scale range; compare $\Pi'^a$ to $\Pi^a$ and
$\Pi_{u'}^a$ in \cref{fig:PI}.

The orientation-averaged inter-scale transfer of total fluctuating
energy by the stochastic fluctuations, $\Pi_{u'}^a$, ceases to be
constant at scales $r$ larger than $0.4d$: indeed, at this point
$(x_{1}, x_{2})=(8d,0)$, $-\Pi_{u'}^a$ is an increasing positive
function of $r$ in the range $0.4d<r<d$, mirroring the decrease of
$-{\Pi_{\tilde{u}}'}^a$ as a function of $r$ towards increasingly
negative values (see \cref{fig:Pi_stochphase_ravg_x28}). These two
contributions add up in \cref{eq:total_Pi_notilde} to give a total
inter-scale transfer rate $\Pi^a$ which is approximately constant over
a range of scales extended well beyond $r=0.4d$, as evidenced in
\cref{fig:Pi_stochphase_ravg_x28}. The correcting action of
$-{\Pi'_{\tilde{u}}}^{a}$ (orientation-averaged energy transfer rate
of stochastic energy by coherent fluctuations), via its positive
values at length-scales $r> 0.4d$, is the essential ingredient for the
extension of the near-constancy of $\Pi^a$ and its near-equality to
$-\varepsilon$ over a range of scales which reaches as far out as
$r=d$. Note that the fully stochastic inter-scale transfer rate
$\Pi'^{a}$ also shows a tendency for being approximately constant over
this range at $(x_{1}, x_{2})=(8d,0)$ (see
\cref{fig:stochKHMH_ravg_x28}) but its values are less close to
$-\varepsilon$ and less constant than
${\Pi}^a_{u'}+{\Pi'_{\tilde{u}}}^a$. The coherent structures play a
definite role in bringing $\Pi^{a}$ closer to a constant equal to
$-\varepsilon$ at larger separations in this near-field flow.

In summary, at both locations $(x_{1}, x_{2})=(2d,0)$ and $(8d,0)$,
the reference equality
\begin{equation}\label{eq:decomposedPiconst}
-{\Pi}^a_{u'}-{\Pi'_{\tilde{u}}}^a \approx \varepsilon^a_r 
\end{equation}
is not too far from our observations in the range where
$\Pi^a\approx\mathrm{const}$. This equality re-writes
$\Pi^a\approx\mathrm{const}$ with more information and this range
increases as one moves downstream along the centreline reaching at
least $\lambda < r < d$ at $(x_{1}, x_{2})=(8d,0)$. We stress that
\cref{eq:decomposedPiconst} is not exactly true. It might be more
accurate to introduce a coefficient multiplying the right hand side
$\varepsilon^a_r$ that is slightly smaller than 1 and not perfectly
constant with $r$; but \cref{eq:decomposedPiconst} is an important
reference formula for our discussion which is not concerned, at this
stage, with exact details. The coherent structures play an important
role in both terms of the left hand side of
\cref{eq:decomposedPiconst} at both locations $(x_{1}, x_{2})=(2d,0)$
and $(8d,0)$, but the stochastic fluctuations do too and more so at
$(x_{1}, x_{2})=(8d,0)$ than $(x_{1}, x_{2})=(2d,0)$.

The approximate balance $\Pi^{a} \approx -\varepsilon$ may be
reminiscent of a Kolmogorov equilibrium cascade but the Kolmogorov
theory is applicable to statistically homogeneous equilibrium
turbulence which is far from the kind of turbulence in the present
near-field wake. This approximate balance follows here from the
approximate balance \cref{eq:decomposedPiconst} and is partly
supported by the effects of the coherent motions on the inter-scale
turbulent energy transfers. The inter-scale transfer rate $\Pi^{a}$
must therefore depend on the inlet/boundary conditions because of the
memory carried by the coherent motions, as it also of course depends
on the kinetic energy and size of the local large scale turbulent
eddies. It is therefore not possible to derive a scaling for $\Pi^{a}$
dimensionally, which means that it is not so easy to use the
approximate balance $\Pi^{a} \approx -\varepsilon$ to derive a scaling
for $\varepsilon$ either. One can derive a scaling for the turbulence
dissipation rate in the context of Kolmogorov equilibrium turbulence
precisely because the inter-scale transfer rate is taken to be
independent of inlet/initial/boundary conditions in this
context. \citet{Goto2016a} have proposed a dissipation balance from
which to derive turbulence dissipation scalings in non-stationary
turbulence with a non-equilibrium cascade, and
\cite{AlvesPortela2018a} have successfully adapted and applied this
balance to the present near-field turbulent wake.

\subsection{Inhomogeneity Contributions to the Non-linear Inter-scale Energy Transfer}\label{subsec:inhomog_scale_transfer}

In interpreting our results, it is relevant to dissociate the
potential contribution of inhomogeneity to the inter-scale transfer
rates. Given that
\begin{equation}\label{eq:hom-condi-1}
\delta {\bf u} \delta q^{2} = {\bf u}^{+} \vert {\bf u}^{+} \vert^{2}
- {\bf u}^{-} \vert {\bf u}^{-} \vert^{2} + {\bf u}^{+} \vert {\bf
  u}^{-} \vert^{2} - {\bf u}^{-} \vert {\bf u}^{+} \vert^{2} -2\delta
{\bf u} ({\bf u}^{-}\cdot{\bf u}^{+})
\end{equation}
one can see that statistical inhomogeneity can make a contribution to
the average of $\delta {\bf u} \delta q^{2}$, at the very least from a
non-zero average of ${\bf u}^{+} \vert {\bf u}^{+} \vert^{2} - {\bf
  u}^{-} \vert {\bf u}^{-} \vert^{2}$. However, we are mainly
concerned with the non-linear inter-scale transfer rate ${\partial
  \over \partial r_{i}}(\delta u_{i} \delta q^{2})$ which has the
property of being $0$ at ${\bf r}=0$ because it is equal to $ \delta
u_{i} {\partial \over \partial r_{i}} \delta q^{2}$ by
incompressibility. We seek a decomposition of ${\partial \over
  \partial r_{i}}(\delta u_{i} \delta q^{2})$ into an inhomogeneity
term and a term unaffected by inhomogeneity such that both vanish at
${\bf r}=0$. Given that ${\partial \over \partial r_{i}} (u^{+}_{i}
\vert {\bf u}^{+} \vert^{2} - u^{-}_{i} \vert {\bf u}^{-}\vert^{2}) =
      {1\over 2}{\partial \over \partial \xi^{+}_{i}} (u^{+}_{i} \vert
      {\bf u}^{+}\vert^{2}) + {1\over 2}{\partial \over \partial
        \xi^{-}_{i}} (u^{-}_{i} \vert {\bf u}^{-}\vert^{2})$ where
      $\xi^{+}_{i}= x_{i} + r_{i}/2$ and $\xi^{-}_{i}= x_{i}
      -r_{i}/2$, it is clear that ${\partial \over \partial r_{i}}
      (u^{+}_{i} \vert {\bf u}^{+} \vert^{2} - u^{-}_{i} \vert {\bf
        u}^{-}\vert^{2})$ is not $0$ at ${\bf r}=0$. We must therefore
      complement the inhomogeneity term ${\partial \over \partial
        r_{i}} (u^{+}_{i} \vert {\bf u}^{+} \vert^{2} - u^{-}_{i}
      \vert {\bf u}^{-}\vert)$ in such a way that the resulting
      inhomogeneity term cancels when ${\bf r}=0$. Starting from
\begin{equation}\label{eq:hom-condi-2}
{\partial \over \partial r_{i}}(\delta u_{i} \delta q^{2}) = {\partial
  \over \partial r_{i}}[\delta u_{i}(\vert {\bf u}^{+} \vert^{2} +
  \vert {\bf u}^{-}\vert^{2})] -2 {\partial \over \partial
  r_{i}}(\delta u_{i} {\bf u}^{-}\cdot {\bf u}^{+})
\end{equation}
it rigorously follows that
\begin{equation}\label{eq:hom-condi-3}
  {\partial \over \partial r_{i}}(\delta u_{i} \delta q^{2}) ={1\over
    2}{\partial \over \partial x_{i}}[u^{+}_{i}\vert {\bf
      u}^{+}\vert^{2} + u^{-}_{i}\vert {\bf u}^{-} \vert^{2} -
    u^{-}_{i}\vert {\bf u}^{+} \vert^{2} - u^{+}_{i}\vert {\bf u}^{-}
    \vert^{2}] - 2{\partial \over \partial r_{i}}(\delta u_{i} {\bf
    u}^{-}\cdot {\bf u}^{+})
\end{equation}
where both the inhomogeneity term ${1\over 2}{\partial \over \partial
  x_{i}}[u^{+}_{i}\vert {\bf u}^{+}\vert^{2} + u^{-}_{i}\vert {\bf
    u}^{-} \vert^{2} - u^{-}_{i}\vert {\bf u}^{+} \vert^{2} -
  u^{+}_{i}\vert {\bf u}^{-} \vert^{2}]$ and the inter-scale transfer
term $- 2{\partial \over \partial r_{i}}(\delta u_{i} {\bf u}^{-}\cdot
{\bf u}^{+})$ vanish at ${\bf r}=0$ (by virtue of incompressibility in
the case of the inter-scale transfer term). The average value of the
inhomogeneity term, $4\Pi_{I} \equiv {1\over 2}{\partial \over
  \partial x_{i}}\langle u^{+}_{i}\vert {\bf u}^{+}\vert^{2} +
u^{-}_{i}\vert {\bf u}^{-} \vert^{2} - u^{-}_{i}\vert {\bf u}^{+}
\vert^{2} - u^{+}_{i}\vert {\bf u}^{-} \vert^{2}\rangle$ can be
non-zero in inhomogeneous turbulence but equals zero in homogeneous
turbulence.  It is clear that $\Pi_{I} =0$ when the turbulence is
statistically homogeneous. Unlike $\Pi_{I}$, the average value of the
pure inter-scale term, $4\Pi_{H}\equiv -2{\partial \over \partial
  r_{i}}\langle \delta u_{i} {\bf u}^{-}\cdot {\bf u}^{+}\rangle$ can
take non-zero values when the turbulence is statistically homogeneous.

We therefore have the decomposition
\begin{equation}\label{eq:hom-condi-4}
 \Pi = \Pi_{I} + \Pi_{H}
\end{equation}
where (i) all three terms ($\Pi$, $\Pi_{I}$ and $\Pi_{H}$) vanish at
${\bf r}=0$, (ii) $\Pi_{I}$ can only be non-zero in the presence of
inhomogeneity and (iii) $\Pi_{H}$ has the exact same form as $\Pi$ in
the case of homogeneous turbulence because $\frac{\partial
  \langle\delta u_i \delta q^2\rangle}{\partial r_i} = -2{\partial
  \over \partial r_{i}}\langle \delta u_{i} {\bf u}^{-}\cdot {\bf
  u}^{+}\rangle$ in such turbulence. This decomposition distinguishes
between a term, $\Pi_{I}$, that is clearly directly accountable to
spatial inhomogeneities, and an inter-scale transfer rate $\Pi_H$
which we may conjecture to be unaffected by spatial
inhomogeneities. In relation to such a conjecture, we must ask whether
our decomposition is unique.

Other such decompositions should take the form
\begin{equation}\label{eq:hom-condi-5}
\Pi = (\Pi_{I} + \Pi_{IH})+ (\Pi_{H} - \Pi_{IH})
\end{equation}
where $\Pi_{IH}$ must meet two conditions: (i) it must equal zero at
${\bf r}=0$ and (ii) it must vanish when the turbulence is
statistically homogeneous. On account of this second condition, we
write $\Pi_{IH} = {\partial \over \partial x_{i}}
\Phi^{x}_{i}$. Because we are dealing with third order statistics we
assume that $\Phi^{x}_{i}$ can only be a sum of products of three
velocity components and the most general way to write this is as
follows:
\begin{multline}\label{eq:hom-condi-6}
\Phi^{x}_{i}=\alpha_{1}\langle u^{+}_{i}\vert {\bf
  u}^{+}\vert^{2}\rangle + \alpha_{2}\langle u^{+}_{i} {\bf
  u}^{-}\cdot {\bf u}^{+}\rangle + \alpha_{3}\langle u^{+}_{i}\vert
    {\bf u}^{-}\vert^{2}\rangle\\ + \beta_{1}\langle u^{-}_{i}\vert
    {\bf u}^{-}\vert^{2}\rangle + \beta_{2}\langle u^{-}_{i} {\bf
      u}^{-}\cdot {\bf u}^{+}\rangle + \beta_{3}\langle u^{-}_{i}\vert
    {\bf u}^{+}\vert^{2}\rangle
\end{multline}
where $\alpha_1$, $\alpha_2$, $\alpha_3$, $\beta_{1}$, $\beta_{2}$,
$\beta_{3}$ are dimensionless constants. With some care it easily
follows that the condition $\Pi_{IH}=0$ for ${\bf r}=0$ implies
\begin{equation}\label{eq:hom-condi-7}
\alpha_{1} + \alpha_{2} + \alpha_{3} + \beta_{1} + \beta_{2} +
\beta_{3} = 0.
\end{equation}

Given that $ \Pi_{IH}$ contributes to the part $(\Pi_{H} - \Pi_{IH})$
of the decomposition, it must be possible to express it in the form
$\Pi_{IH} = {\partial \over \partial r_{i}} \Phi^{r}_{i}$. To find the
conditions for this to be possible, we use ${\partial \over \partial
  x_{i}}= {\partial \over \partial \xi^{+}_{i}} + {\partial
  \over \partial \xi^{-}_{i}}$ and use $\Pi_{IH} = {\partial \over
  \partial x_{i}} \Phi^{x}_{i}$ to write
\begin{multline}\label{eq:hom-condi-8}
\Pi_{IH}=\alpha_{1}{\partial \over \partial \xi^{+}_{i}}\langle
u^{+}_{i}\vert {\bf u}^{+}\vert^{2}\rangle + \alpha_{2}{\partial \over
  \partial \xi^{+}_{i}}\langle u^{+}_{i} {\bf u}^{-}\cdot {\bf
  u}^{+}\rangle + \beta_{2}{\partial \over \partial
  \xi^{+}_{i}}\langle u^{-}_{i} {\bf u}^{-}\cdot {\bf u}^{+}\rangle +
\beta_{3}{\partial \over \partial \xi^{+}_{i}} \langle u^{-}_{i}\vert
     {\bf u}^{+}\vert^{2}\rangle\\+\alpha_{2}{\partial \over \partial
       \xi^{-}_{i}}\langle u^{+}_{i} {\bf u}^{-}\cdot {\bf
       u}^{+}\rangle + \alpha_{3}{\partial \over \partial
       \xi^{-}_{i}}\langle u^{+}_{i}\vert {\bf u}^{-}\vert^{2}\rangle
     + \beta_{1}{\partial \over \partial \xi^{-}_{i}}\langle
     u^{-}_{i}\vert {\bf u}^{-}\vert^{2}\rangle + \beta_{2} {\partial
       \over \partial \xi^{-}_{i}} \langle u^{-}_{i} {\bf u}^{-}\cdot
     {\bf u}^{+}\rangle .
\end{multline}
Note that $\alpha_{1}{\partial \over \partial \xi^{+}_{i}}\langle
u^{+}_{i}\vert {\bf u}^{+}\vert^{2}\rangle = 2\alpha_{1}{\partial
  \over \partial r_{i}}\langle u^{+}_{i}\vert {\bf
  u}^{+}\vert^{2}\rangle$ because ${\partial \over \partial
  \xi^{-}_{i}}\langle u^{+}_{i}\vert {\bf u}^{+}\vert^{2}\rangle = 0$
and ${\partial \over \partial r_{i}}= {1\over 2}({\partial \over
  \partial \xi^{+}_{i}} - {\partial \over \partial \xi^{-}_{i}})$. For
the same reason, $\beta_{1}{\partial \over \partial
  \xi^{-}_{i}}\langle u^{-}_{i}\vert {\bf u}^{-}\vert^{2}\rangle =
-2\beta_{1}{\partial \over \partial r_{i}}\langle u^{-}_{i}\vert {\bf
  u}^{-}\vert^{2}\rangle$. All the other terms and combinations of
other terms cannot be rephrased in ${\partial \over \partial r_{i}}$
form. The necessary form $\Pi_{IH} = {\partial \over \partial r_{i}}
\Phi^{r}_{i}$ then implies $\alpha_{2} = \alpha_{3}=
\beta_{2}=\beta_{3}=0$. From \cref{eq:hom-condi-7} follows $\alpha_{1}
= -\beta_{1}$ and therefore
\begin{equation}\label{eq:hom-condi-9}
\Pi_{IH}= \alpha {\partial \over \partial r_{i}}(\langle
u^{+}_{i}\vert {\bf u}^{+}\vert^{2}\rangle +\langle u^{-}_{i}\vert
{\bf u}^{-}\vert^{2}\rangle) = \alpha {\partial \over \partial
  x_{i}}(\langle u^{+}_{i}\vert {\bf u}^{+}\vert^{2}\rangle -\langle
u^{-}_{i}\vert {\bf u}^{-}\vert^{2}\rangle).
\end{equation}
where we also made use of $\Pi_{IH} = {\partial \over \partial x_{i}}
\Phi^{x}_{i}$ with \cref{eq:hom-condi-6} and \cref{eq:hom-condi-7},
and where we set $\alpha \equiv 2\alpha_{1}$.

In conclusion, the decomposition \cref{eq:hom-condi-4} is not unique
as one can always use $\Pi_{IH}$ given by \cref{eq:hom-condi-9} to
obtain another equally valid decomposition
\cref{eq:hom-condi-5}. However, if one averages over scale-space
orientations, the decomposition
\begin{equation}\label{eq:hom-condi-10}
\Pi^{a} = \Pi_{I}^{a} + \Pi_{H}^{a}
\end{equation}
is unique because $\Pi_{IH}^{a}=0$ given that $\Pi_{IH}$ in
\cref{eq:hom-condi-9} is such that $\Pi_{IH} ({\bf r}) = - \Pi_{IH}
(-{\bf r})$. The conjecture that the orientation-averaged inter-scale
transfer rate $\Pi_{H}^{a}$ may be unaffected by spatial
inhomogeneities is more likely to hold than the conjecture that
$\Pi_{H}$ is unaffected by spatial inhomogeneities. This conjecture
and the decomposition introduced in this subsection are an attempt at
introducing a tool which can help make some analytic sense of the
concept of an inhomogeneous turbulence cascade.

\begin{figure}
    \centering
    \begin{subfigure}[b]{\textwidth}
        \centering
                \includegraphics{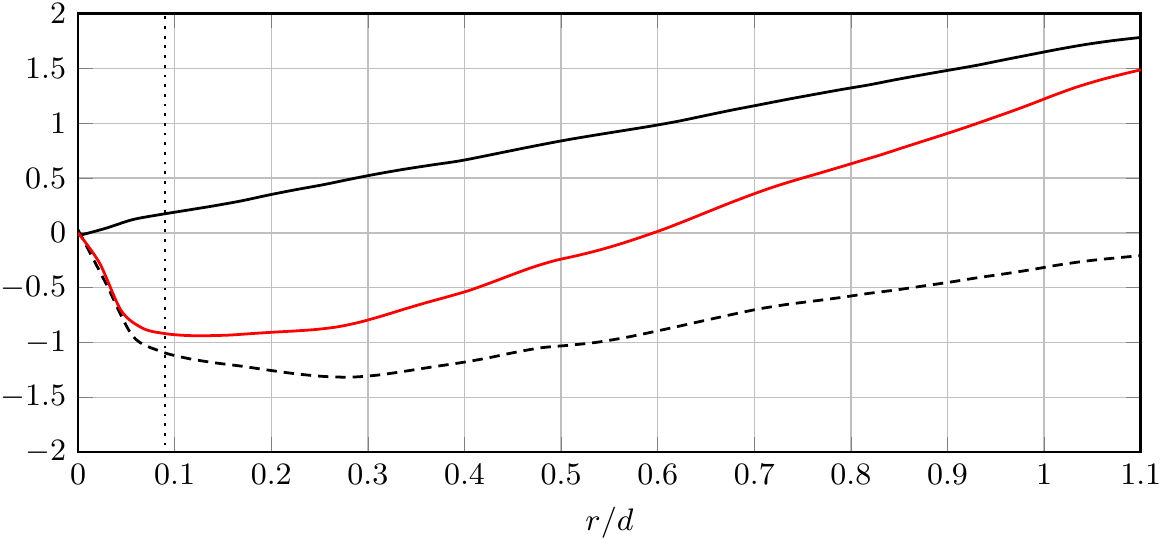}
    \end{subfigure}%
\\
    \begin{subfigure}[b]{\textwidth}
        \centering
               \includegraphics{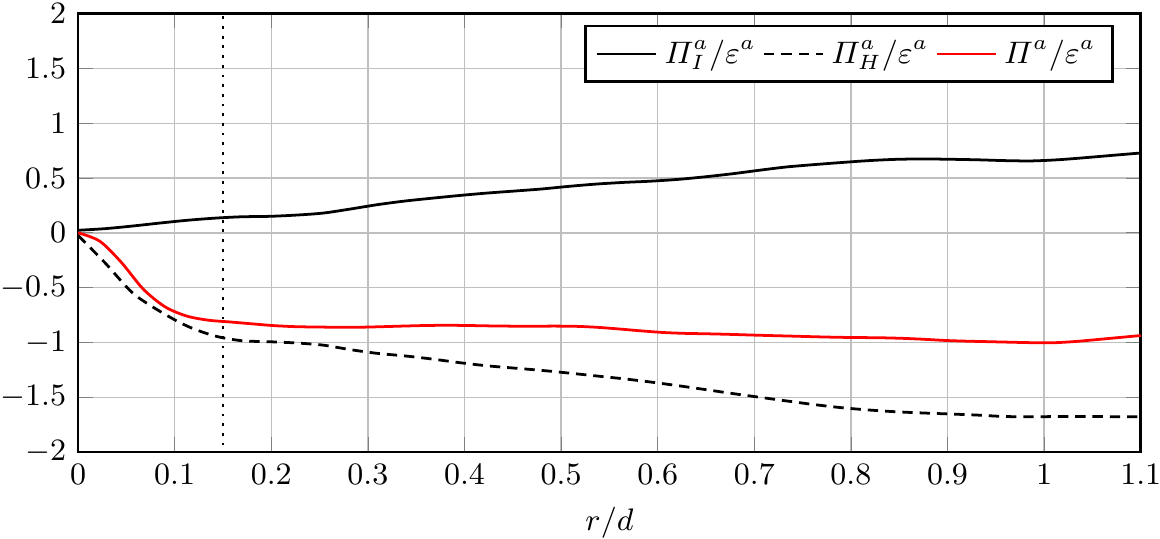}
    \end{subfigure}
    \caption{Orientation averaged inter-scale energy transfer terms
      $\Pi_{I}^{a}$, $\Pi_{H}^{a}$ and $\Pi^{a}$ (see equations
      \ref{eq:hom-condi-3} and \ref{eq:hom-condi-4}) at $x_1/d=2$
      (top) and $x_1/d=8$ (bottom). The vertical dotted line gives the
      position of $r=\lambda$.}\label{fig:Pi_hominhom}
\end{figure}

In \cref{fig:Pi_hominhom} we plot the orientation averaged inter-scale
transfer rates $\Pi^{a}$, $\Pi_{I}^{a}$ and $\Pi_{H}^{a}$ at the two
centreline positions $(x_{1}, x_{2})=(2d,0)$ and
$(8d,0)$. Inhomogeneity inter-scale transfer is present and positive at
all scales, but may be considered negligible at dissipative scales $r$
smaller than $0.1d$, i.e. smaller than the Taylor microscale
$\lambda$. However, it does make a significant contribution to the
total inter-scale energy transfer rate $\Pi^{a}$ at scales $r$ larger
than $\lambda$, particularly at $(x_{1}, x_{2})=(2d,0)$ where
$\Pi_{I}^{a}$ is commensurate throughout these scales with the
negative inter-scale transfer $\Pi_{H}^{a}$. In fact $\Pi^{a}$ changes
sign from negative to positive as $r$ increases beyond $r\approx 0.6d$
because of the influence of the positive inhomogeneity inter-scale
energy transfer rate.

In \cref{fig:Pi_hominhom} one can also see that the contribution of
the inhomogeneity part of the inter-scale energy transfer weakens with
downstream distance, while remaining positive throughout the scales.
$\Pi^{a}$ and $\Pi_{H}^{a}$ are both negative throughout the scales
and significantly closer to each other than to $\Pi_{I}^{a}$ at
$(x_{1}, x_{2})=(8d,0)$, which is not the case at $(x_{1},
x_{2})=(2d,0)$.

It is particularly intriguing that $\Pi^{a}$ would not have been
approximately constant across the scales, from about $\lambda$ to
about $0.3d$ at $(x_{1}, x_{2})=(2d,0)$ and from about $\lambda$ to
about $d$ at $(x_{1}, x_{2})=(8d,0)$, without the inhomogeneity
contribution coming from $\Pi_{I}^{a}$. It is in fact this
inhomogeneity contribution which returns a near-constancy of $\Pi^{a}$
all the way up to scales $r$ equal to $d$ at $(x_{1}, x_{2})=(8d,0)$
and imparts on the orientation-averaged inter-scale energy transfer
$\Pi^{a}$ a Kolmogorov-seeming behaviour over a decade of scales $r$.

The results of these two subsections suggest that the approximate
balance $\Pi^{a} \approx -\varepsilon$ observed in our turbulent
wake's very near field, even if reminiscent of a Kolmogorov
equilibrium for homogeneous turbulence, is in fact possible in this
near-field turbulence because of the presence of spatial inhomogeneity
and coherent structures.

\subsection{Inter-scale fluxes}\label{subsubsec:fluxes}

In order to interpret the inter-scale physics behind the negative sign
of $\Pi^{a}$ it is necessary to also consider the inter-scale flux
$\langle \delta {\bf u} \delta q^{2} \rangle$ given that $\Pi$ is the
divergence of this flux in scale space ${\bf r}$. In particular, it is
necessary to consider the sign of the radial component of the
orientation-averaged inter-scale flux. One cannot claim that the
inter-scale energy transfer proceeds from large to small scales on
average if this sign is not negative too.

The inter-scale flux vectors which correspond to each term in
\cref{eq:total_Pi_tmp} are related by
\begin{equation}\label{eq:total_flux_tmp}
  \langle \delta {\bf u} \delta q^{2} \rangle = \langle \delta {\bf
    u}' \delta q'^{2} \rangle + \langle \delta {\bf \tilde{ u}} \delta
  q'^{2} \rangle + \langle \delta {\bf \tilde{ u}} \delta
  \tilde{q}^{2} \rangle + 2\langle \delta {\bf u}' (\delta{\bf
    u}'\cdot\delta{\bf \tilde{ u}})\rangle .
\end{equation}
The flux vectors are placed in this equation in exactly the same way
as their corresponding inter-scale transfer rates are placed in
\cref{eq:total_Pi_tmp}. The inter-scale flux identity which reflects
$\tilde{\Pi}_{\mathcal{P}_{\tilde{u}}} = \Pi_{u'} - \Pi'$ is $2\langle
\delta {\bf u}' (\delta{\bf u}'\cdot\delta{\bf \tilde{ u}})\rangle =
\langle \delta {\bf u}' \delta q^{2} \rangle - \langle \delta {\bf u}'
\delta q'^{2} \rangle$. Combined with \cref{eq:total_flux_tmp} it
yields
\begin{equation}\label{eq:total_flux}
  \langle \delta {\bf u} \delta q^{2} \rangle = \langle \delta {\bf
    u}' \delta q^{2} \rangle + \langle \delta {\bf \tilde{ u}} \delta
  q'^{2} \rangle + \langle \delta {\bf \tilde{ u}} \delta
  \tilde{q}^{2} \rangle 
\end{equation}
which corresponds to \cref{eq:total_Pi}.

\begin{figure}
    \centering
    \begin{subfigure}[b]{\textwidth}
        \centering
        \includegraphics{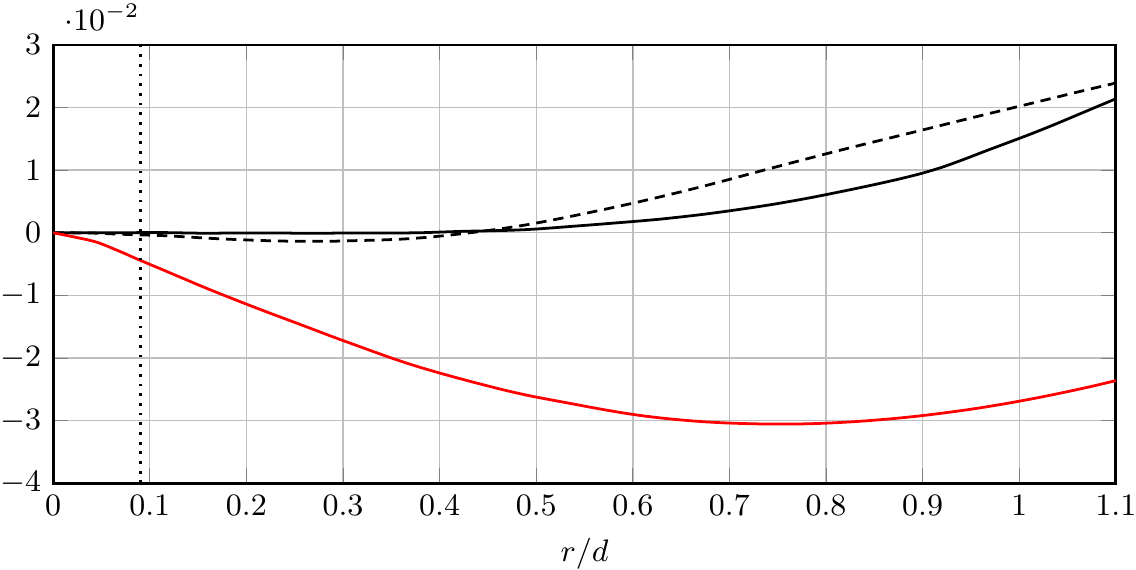}
    \end{subfigure}%
\\
    \begin{subfigure}[b]{\textwidth}
        \centering
        \includegraphics{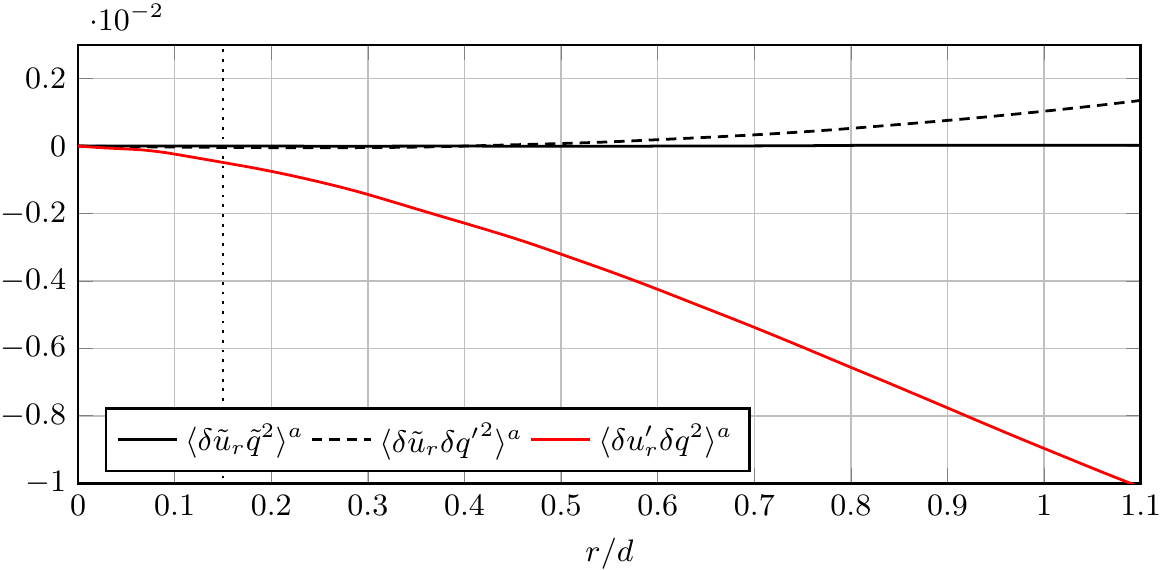}
    \end{subfigure}
    \caption{Orientation averaged non-linear inter-scale radial fluxes
      terms at $x_1/d=2$ (top) and $x_1/d=8$ (bottom). The vertical
      dotted line gives the position of
      $r=\lambda$.}\label{fig:dudq_ravg_x28}
\end{figure}

We are interested in the orientation-averaged radial components of
these fluxes in the $r_{3}=0$ plane. In \cref{fig:dudq_ravg_x28} we
plot, as functions of $r$, the orientation-averaged radial components
(in the $r_{3}=0$ plane) $\langle\delta{u'}_r\delta{q}^2\rangle^a$,
$\langle\delta\tilde{u}_r\delta{q'}^2\rangle^a$ and
$\langle\delta\tilde{u}_r\delta\tilde{q}^2\rangle^a$. The latter is
zero where \cref{eq:decomposedPiconst} is relevant. Concentrating our
attention on the scale range where \cref{eq:decomposedPiconst} is
relevant, the signs of these orientation-averaged radial fluxes and of
the corresponding orientation-averaged inter-scale transfer rates
therefore suggest the following: (i) concerning $\Pi_{u'}^a$, the
stochastic fluctuations transfer, on average, total (stochastic and
coherent) fluctuating energy from large to small scales in the range
$r<0.3d$ at $(x_{1}, x_{2})=(2d,0)$ and $r<d$ at $(x_{1},
x_{2})=(8d,0)$; (ii) concerning ${\Pi'_{\tilde{u}}}^a$, the coherent
fluctuations transfer, on average, stochastic energy from large to
small scales at length-scales $r<0.3d$ at both spatial locations, but
from small to large scales at $(x_{1}, x_{2})=(8d,0)$ in the range
$0.4d<r<d$. The contribution of ${\Pi'_{\tilde{u}}}^{a}$ is the
smallest of the two inter-scale transfer rate terms, $\Pi_{u'}^a$ and
${\Pi'_{\tilde{u}}}^{a}$, in \cref{eq:total_Pi_notilde} and
\cref{eq:decomposedPiconst}. The inter-scale fluctuating energy
transfer proceeds, therefore, from large to small scales on average,
mostly because of the large to small scale transfer of total
fluctuating energy by stochastic fluctuations.

\section{Conclusions}\label{sec:conclusion}

By conditionally sampling the fluctuating velocity and pressure fields
in the wake generated by a square prism (as introduced in the
classical work of \citet{Hussain1970}), those fluctuating fields were
decomposed into two components: a phase averaged component whose time
signature follows the vortex shedding and a stochastic component which
can be interpreted as the turbulent fluctuations which are
superimposed onto the organised motion associated with the vortex
shedding. Taking also into account the corresponding mean fields, we
used the inter-scale and inter-space energy balance, the KHMH
equation, written for a triple decomposition and we analysed DNS data
of a near-field turbulent wake. Our study has been limited to the
geometric centreline and the plane of the mean flow. The turbulence in
this near wake, at a distance between $2d$ and $8d$ of the square
prism, is very inhomogeneous and very unsteady. Unsurprisingly, the
non-stationarity and inhomogeneity contributions to the KHMH balance
dominate. The pressure-velocity term is sizeable too, particularly at
scales $r$ larger than about $0.4d$, and has an orientation signature
which appears similar to that of the purely stochastic non-linear
inter-scale transfer rate.

We reduced the amount of information by taking orientation averages of
every term in the KHMH equation. In an orientation-averaged sense, the
production of kinetic energy by the mean flow does not feed the
stochastic turbulent fluctuations directly. Instead, energy is
transferred from the mean flow to the coherent fluctuations which in
turn transfer energy to the stochastic fluctuations. The coherent
structures also dominate spatial turbulent transport of small-scale
two-point stochastic turbulent fluctuations.

\citet{AlvesPortela2017} found that the orientation-averaged
non-linear inter-scale transfer rate $\Pi^{a}$ is approximately
independent of $r$ in the scale-ranges $\lambda\le r \le 0.3d$ and
$\lambda\le r\le d$, respectively, at stream-wise distances $x_{1}=2d$
and $x_{1}=8d$ from the square prism. We have shown here that this
requires a definite inter-scale transfer contribution by the coherent
structures at $x_{1}=2d$ but not at $x_{1}=8d$ where it is mostly
attributable to stochastic fluctuations. However, at $x_{1}=8d$,
$-\Pi^{a}$ is also very close to $\varepsilon$ in the range
$\lambda\le r\le d$ and the contribution of the coherent structure's
inter-scale energy transfer is a significant factor in achieving this
approximate equality. The later contribution, albeit relatively small,
appears to resist the energy transfer in the direct sense since
${\Pi'_{\tilde{u}}}^a>0$ at large enough scales. The self-interaction
of the coherent motions plays a negligible role in the inter-scale
energy transfer.

The inter-scale energy transfer rate can be decomposed in two terms,
one which is absent in homogeneous turbulence and therefore relates
directly to spatial inhomogeneity, and another which remains present
in homogeneous turbulence. One might be able to consider the concept
of inhomogeneity-induced inter-scale energy transfers alongside the
usual homogeneous inter-scale energy transfers. Perhaps most
surprisingly and most importantly, a very significant direct
contribution to the inter-scale energy transfer rate turns out to come
from spatial inhomogeneity without which the approximate equality
$-\Pi^{a} \approx \varepsilon$ would not have been possible in this
very near field..

\section*{Acknowledgements}
The authors acknowledge the EU support through the FP7 Marie Curie
MULTISOLVE project (grant no. 317269) as well as the computational
resources allocated in ARCHER HPC through the UKTC funded by the EPSRC
grant no. EP/L000261/1. JCV also acknowledges the support of an ERC
Advanced Grant (grant no. 320560) and Chair of Excellence CoPreFlo
funded by I-SITE/MEL/Region Hauts de France. Declaration of interests.
The authors report no conflicts of interest.

\FloatBarrier
\bibliographystyle{template/jfm}

\bibliography{library}

\end{document}